\documentclass
[11pt,tightenlines,eqsecnum,floats,aps,amsmath,amssymb,nofootinbib,prd,shownopacs,notitlepage]{revtex4}
\usepackage{graphicx, wrapfig}
\usepackage{amssymb}
\usepackage{color}
\usepackage{mathrsfs}
\usepackage{amsmath}
\usepackage{subfigure}
\usepackage{afterpage}

\usepackage[colorinlistoftodos]{todonotes}
\usepackage{epstopdf}

\DeclareGraphicsExtensions{.pdf,.png,.jpg,.jpeg,.eps}

\def\f{\frac}

\def\rhopl{\rho_{\rm Pl}}

\def\heff{\mathcal{H}_{\rm eff}}
\def\heffgen{\mathcal{H}_{\rm eff}^{\rm gen}}

\def\b{\bar}
\def\lp{\ell_{\rm Pl}}

\def\t{\tilde}
\def\h{\hat}
\def\R{\mathcal{R}}

\def\ag{{\Delta_{o}}}
\def\EE {\rm EEs\,}
\def\GEE{\rm GEEs\,\,}
\def\Hp{\mathcal{H}_{\rm phy}}
\def\Gammaq{\Gamma_{\rm Q}}
\def\Gammac{\Gamma_{\rm C}}
\def\Xq{X_{\rm Q}}

\def\rhosup{\rho_{\rm sup}}
\def\rhob{\rho_{{}_{\rm B}}}

\def\l{\ell_{o}}
\def\rmd{{\rm d}}
\def\ab{\bar{a}}
\def\gb{\bar{g}}
\def\pb{\bar{p}}
\def\Vb{\overline{V}}
\def\tb{\bar{t}}

\usepackage{enumerate}

\newcommand{\be}{\nopagebreak[3]\begin{equation}}
\newcommand{\ee}{\end{equation}}
\newcommand{\bfig}{\nopagebreak[3]\begin{figure}}
\newcommand{\efig}{\end{figure}}
\newcommand{\ba}{\nopagebreak[3]\begin{eqnarray}}
\newcommand{\ea}{\end{eqnarray}}

\newcommand{\bmult}{\nopagebreak[3]\begin{multline}}
\newcommand{\emult}{\end{multline}}

\begin{document}
\preprint{IGC-15/9-2}
\title{Generalized effective description of loop quantum cosmology}
\author{Abhay Ashtekar}
\email{ashtekar@gravity.psu.edu} 
\author{Brajesh Gupt}
\email{bgupt@gravity.psu.edu}
\affiliation{
Institute for Gravitation and the Cosmos \& Physics Department, The Pennsylvania State University, University Park, PA 16802 U.S.A.
}


\begin{abstract}

The effective description of loop quantum cosmology (LQC) has proved to be a convenient platform to study phenomenological implications of the quantum bounce that resolves the classical big-bang singularity. Originally, this description was derived using Gaussian quantum states with small dispersions. In this paper we present a generalization to incorporate states with large dispersions. Specifically, we derive the \emph{generalized} effective Friedmann and Raychaudhuri equations and propose a generalized effective Hamiltonian which are being used in an ongoing study of the phenomenological consequences of a broad class of quantum geometries. We also discuss an interesting interplay between the physics of states with larger dispersions in standard LQC, and of sharply peaked states in (hypothetical) LQC theories with larger area gap.
\end{abstract}

\maketitle

\section{Introduction}
\label{s1}

Loop quantum gravity (LQG) has led to a specific quantum geometry \cite{alrev,crbook,ttbook} that replaces the Riemannian geometry used in all modern gravitational theories. The primary goal of loop quantum cosmology (LQC) is to investigate the implications of the quantum nature of this geometry in the cosmological sector of general relativity. Specifically, in LQG one can naturally define self-adjoint operators representing geometric observables --such as areas of physical surfaces and volumes of physical regions \cite{al5,al-vol}. Riemannian geometry is quantized in the direct sense that eigenvalues of these geometric operators are discrete \cite{rs,al5,al-vol}. In particular, there is a smallest non-zero eigenvalue of the area operator, which represents the fundamental \emph{area gap} of the theory, denoted $\ag$ (in Planck units) \cite{al5}. This \emph{microscopic} parameter sets the scale for new phenomena that distinguish LQC from both general relativity and the Wheeler-DeWitt theory. In particular, dynamics of general relativity is drastically modified at this scale, leading to a natural resolution of the big-bang singularity in a variety of cosmological models. These include the k=0 and k=1 (FLRW) models \cite{mb1,abl,aps1,aps2,aps3,acs,ps,apsv,warsaw1}, possibly with a non-zero cosmological constant \cite{bp,kp1,ap}, the anisotropic Bianchi I, II and IX models \cite{awe2,madrid-bianchi,awe3,we} and the simplest of the inhomogeneous models ---the Gowdy space-times--- widely studied in exact general relativity \cite {hybrid1,hybrid2,hybrid3,hybrid4,hybrid5}. (For a review, see, e.g., \cite{asrev}). In this paper we will restrict ourselves to the simplest case, the k=0, Friedmann, Lema\^{i}tre, Robertson, Walker (FLRW) model.

The analysis leading to a natural resolution of the big bang singularity was originally driven by the central conceptual and mathematical questions of quantum gravity. However, in recent years singularity resolution has also had phenomenological applications because it provides a natural avenue to directly address the trans-Planckian issues that arise in the pre-inflationary phase of the dynamics of cosmological perturbations \cite{aan1,aan2,aan3,abrev}. In these and other applications of the LQC singularity resolution, effective equations (\EE) \cite{jw,vt,asrev} have played an important role in the intermediate steps of the analysis because they capture the leading order quantum effects that are responsible for the quantum bounce. However, in the detailed investigations carried out so far, these equations are derived starting from sharply peaked quantum states $\Psi$. While the restriction to sharply peaked states in this analysis is well motivated \cite{aps1,aps3,ap}, in order to probe the robustness of these results, it is of interest to allow more general states and investigate the resulting dynamics. Recently, thanks to a significant improvement in numerical techniques used in LQC, detailed simulations could be performed using widely spread Gaussian and non-Gaussian states $\Psi$ in \emph{full} LQC \cite{dgs2,dgms}. They showed that, in spite of this significant increase in generality, the quantum evolution of states $\Psi$ remains qualitative similar and, in particular, the bounce persists (as expected from the fact that energy density operator has an upper bound $\rhosup$). But, whereas for sharply peaked states the standard \EE track the expectation values of physical observables very accurately \cite{aps3,asrev}, this ceases to be the case for the widely spread states $\Psi$. This is not surprising, given that the derivation of the standard \EE is tailored to sharply peaked quantum states \cite{jw,vt}. 

The goal of this paper is to obtain \emph{generalized} effective equations (\GEE) which do track the expectation values of the principal physical observables accurately even for widely spread states. The groundwork laid down in this paper is being used to analyze how the observable predictions such as the power spectrum of cosmological perturbations change if the underlying wave function $\Psi$ of the background FLRW quantum geometry is not sharply peaked \cite{aag}.

The paper is organized as follows. In section \ref{s2} we recall properties of
the solvable k=0 FLRW model \cite{acs} that are needed in the derivation of the
GEEs. The derivation is presented in section \ref{s3}. These \GEE only track the evolution of expectation values of the principal observables associated with the background FLRW geometry, and not their uncertainties or higher fluctuations. Therefore we will not need the machinery of geometric quantum mechanics that was used in the `embedding method' of deriving the standard \EE in LQC \cite{vt,asrev}. Rather, we will use just the quantum equations governing this model, summarized in section \ref{s2}. Various features of the \GEE are discussed in section \ref{s4}. In particular, there is a 1-1 correspondence between the \GEE associated with widely spread states in LQC, and the standard \EE associated with sharply peaked states in a hypothetical LQC theory with a larger area gap. Thus, the \GEE could also have been derived using geometric quantum mechanics \cite{aats} and the `embedding method' of approximating the quantum dynamics of sharply peaked states \cite{vt,asrev} in a mathematical theory which is identical to LQC but has a larger area gap $\Delta$ than the $\ag$ used in LQC. Section \ref{s4} will also feature a comparison between predictions of \GEE and exact numerical results that underlie the construction of the quantum corrected, dressed metric $\t{g}_{ab}$, used in the study of cosmological perturbations during the pre-inflationary phase. This discussion will show that the differences are typically smaller than the observational errors in CMB measurements. Therefore it is possible to use the \GEE also in the computation of the inflationary power spectrum --and in the investigation of observable effects of pre-inflationary dynamics-- for a large class of widely spread states $\Psi$. Section \ref{s5} summarizes the results and puts them in a broader perspective. Technical intermediate steps used in section \ref{s4} are spelled out in Appendix A.

 \section{Preliminaries}
\label{s2}

This section is divided in two parts. In the first we recall the LQC description of a soluble FLRW model and in the second we discuss the effective description that captures the key quantum features associated with the dynamics of sharply peaked physical states $\Psi$. We have taken this opportunity to rearrange the material by eliminating the Barbero-Immirzi mathematical parameter $\gamma$ in favor of the area gap $\ag$ which plays a direct physical role in the theory. This discussion will also serve to fix the notation and provide the key equations that are used in subsequent analysis.

\subsection{LQC of a soluble model}
\label{s2.1}

In this paper we focus on the k=0 FLRW model with a massless scalar field $\phi$ as the source because this model is exactly solvable both in general relativity and LQC. The analytical results obtained here provide guidance for more realistic inflationary models that are being investigated numerically \cite{aag}. 

As in much of the cosmology literature, we will assume $\mathbb{R}^{3}$ spatial topology. In this case, one encounters trivial infrared divergences already in the classical phase space framework because of homogeneity and spatial non-compactness. The simplest way to address this issue is to work with a fiducial cell $\mathcal{C}$, construct the Hamiltonian framework, pass to quantum theory and remove this infrared cut-off at the end (see, e.g., \cite{asrev}). In LQC, one uses the pair $(v,\phi)$ as configuration variables where $v$ determines the physical volume of the cell $\mathcal{C}$ via%
\footnote{While in geometrodynamics one works with 3-metrics, in LQG one works with their `square-roots', the triads. Therefore the configuration variable $v$ ranges over the entire real line, assuming positive values if the physical triad has the same orientation as the fiducial, co-moving one and negative values if the orientation is opposite. The fact that $v$ is not constrained to be positive  simplifies the quantization procedure. The $2\pi G$ factor is introduced just to simplify the subsequent equations in quantum theory.}
\be V = 2\pi G |v|\, . \ee
In place of the canonically conjugate momentum $p_{v}$, we will use 
$h = 2p_{v}$ because $h$ has a simple space-time interpretation: it equals the Hubble parameter, $\dot{a}/a$, on solutions to classical Einstein's equations. Therefore, the non-vanishing Poisson brackets are given by
\be \{v,\, h\} \, =\, 2 \quad {\rm and} \quad 
\{\phi,\, p_{\phi}\} \, = \, 1.  \ee
This phase space inherits a single Hamiltonian constraint from general relativity. Adapted to the cosmic time $t$ (or, equivalently to the lapse function $N=1$), it reads
\be \label{chc} \mathcal{H}_{\rm GR} := \f{p_{\phi}^{2}}{4\pi G v}\, - \, \f{3  h^{2} v}{4} \approx 0\, . \ee 
This constraint is equivalent to the Friedmann equation and dictates the dynamics of general relativity in the Hamiltonian framework. It turns out that in the quantum theory one can `deparametrize' the quantum Hamiltonian constraint readily if one uses, in place of $t$, a harmonic time variable $\tau$ satisfying $\Box \tau =0$ (for which the lapse is $N_{\tau} = 4\pi G v$). Then the Hamiltonian constraint is rescaled as follows:
\be \label{chc2}   p_{\phi}^{2} - 3\pi G\, (h v)^{2} \, \approx\, 0\, , \ee

To pass to quantum theory, one uses the LQG techniques. While this construction has been carried out both in the $(v, \phi)$ and $(h, \phi)$ representations, exact solubility of the model becomes manifest in the latter \cite{acs} (see Appendix A). In this representation, the quantum analog of (\ref{chc2}) turns out to have the form:%
\footnote{For details on this point and on the results reported in the rest of this section, see \cite{acs} and \cite{asrev}. Our variables $h,\l$ are related to the variables ${\rm b},\lambda$ used there via $h= {\rm b}/\gamma$ and $\l = \gamma\, \lambda$, where $\gamma$ is the Barbero-Immirzi parameter of LQG. In the earlier references both the area gap $\ag$ and the Barbero Immirzi parameter $\gamma$ appear explicitly in key equations, e.g., the expression of $\rhosup$, although the two are in fact linearly related. This obscures the true role played by the area gap itself. The use of $(h, \l)$ removes this redundancy. We will write all equations using only the area gap.}
\be
\label{hc1}\partial_\phi^2 \Psi(h,\,\phi)\, = \, 12\pi G
\Big(\f{\sin(\l h)}{\l}\partial_{h}\Big)^2 \Psi(h,\,\phi)\, , \quad {\rm where}\quad \l^{2} = \f{\Delta_{o}^{3}}{48\pi^{2}}\, \lp^{2}\, .
\ee
Thus, in the quantization based on LQG, because of the underlying quantum geometry, the phase space variable $h$ in (\ref{chc2}) is effectively replaced by the multiplicative operator \,$(\sin \l h)/\l$\, in (\ref{hc1}) in the $(h,\phi)$ representation. Since $h$ is the Hubble parameter on any dynamical trajectory of classical general relativity, it ranges over $(0, \infty)$. 
By contrast, in LQC it ranges over only a finite interval, $(0, \pi/\l)$; the classical range is recovered only in the limit in which the area gap $\ag$ tends to zero. Every physical quantum state $\Psi(h,\phi)$ satisfies (\ref{hc1}) in LQC. 

Note that because the left side of (\ref{hc1}) has two derivatives with respect to $\phi$ and the operator on the right side is independent of $\phi$, (\ref{hc1}) has the form of the Klein-Gordon equation in a (fictitious) static space-time in which $\phi$ plays the role of time and $h$ of the spatial coordinate, with the operator on the right side serving as the spatial Laplacian. In this sense, there is a natural deparametrization of the Hamiltonian constraint in which the dynamical variable $\phi$ serves as time. Using this deparametrization, one can endow the space of physical states $\Psi(h,\phi)$ with an appropriate Hermitian inner product by using a standard group averaging procedure \cite{aps2}. This structure arises naturally once we adapt the classical Hamiltonian constraint (\ref{chc2}) to harmonic time because $\phi$ satisfies $\Box \phi =0$. 

Conceptually, $\phi$ serves as the `internal' or `relational' time variable with respect to which physical observables such as the volume and the matter density evolove. Since the right side of (\ref{hc1}) is $\phi$-independent, $\h{p}_{\phi} = -i\hbar \partial/\partial \phi$ is a constant of motion, and therefore a Dirac observable. The second family of useful Dirac observables is provided by $\h{V}|_{\phi}$, the operators corresponding to the physical volume of the fiducial cell $\mathcal{C}$ at any given time instant $\phi$. The expectation values of these observables turn out to have a surprisingly simple form \cite{acs}:
\be
 \label{V} \langle \widehat{V}\rangle\mid_\phi \,=\, V_{+} e^{\alpha \phi} + V_{-} e^{-\alpha \phi}, \quad {\rm with} \quad \alpha=\sqrt{12\pi G}
\ee
where $V_\pm$ are constants defined by the given physical state $\Psi(h,\phi)$ (see Appendix A). Note that this evolution of $\langle \widehat{V}\rangle\mid_\phi$  is exact and holds for \emph{all} states $\Psi(h,\phi)$ (in the domains of the volume operator); there is no restriction to sharply peaked states. Eq. (\ref{V}) explicitly shows that the universe expands for large positive values of the internal time $\phi$ and contracts for large negative values, for all physical states $\Psi(h,\phi)$. The expectation values of $\h{p}_{\phi}$ are, of course, independent of the internal time variable $\phi$.  

\subsection{The standard effective description}
 \label{s2.2}

In this sub-section, we will restrict ourselves to physical states $\Psi(h,\phi)$ which are \emph{sharply peaked} at a late time $\phi_{o}$ \emph{in the sense that} the relative uncertainties in the Dirac observables $\h{p}_{\phi}$ and $\h{V}|_{\phi_{o}}$ are small. A natural question is whether they continue to remain sharply peaked in these two observables at all times, in particular in the Planck regime. Detailed investigations of the quantum evolution of (\ref{hc1}) showed that the answer is in the affirmative. (While this result may seem surprising at first, it can be understood by considering simpler systems whose Hamiltonian is related to the two observables of interest in a similar way.) It turns out that for these sharply peaked states $\Psi$, using ideas from geometric quantum mechanics \cite{aats} one can systematically derive certain \EE governing the evolution of $v$ and $h$, that provide the key quantum corrections to classical dynamics \cite{jw,vt,asrev}. 

The geometric quantum mechanics framework allows one to regard the space of all quantum states as an infinite dimensional phase space $\Gammaq$ in which the \emph{exact} quantum dynamics is represented by the flow generated by a Hamiltonian vector field $\Xq$, just as in classical mechanics. Furthermore, there is natural projection from $\Gammaq$ to the classical phase space $\Gammac$. The strategy is to obtain quantum corrections by embedding $\Gammac$ into $\Gammaq$. The image $\Gamma$ of this imbedding is of course a sub-manifold of $\Gammaq$ that is naturally isomorphic to $\Gammac$. The idea is to choose $\Gamma$ judiciously so that the Hamiltonian vector field $\Xq$ (implementing the \emph{full quantum} dynamics) is tangential to  $\Gamma$ to a high degree of approximation. In one succeeds in finding such a $\Gamma \subset \Gammaq$, then the projection of $\Xq$ into $\Gamma$ provides a dynamical flow on $\Gamma$ which approximates the full quantum dynamics very well. This flow can be faithfully projected to the classical phase space $\Gammac$ because by construction $\Gamma$ is isomorphic to $\Gammac$, providing us with the desired quantum corrections to classical dynamics. To summarize, the embedding provides a set of quantum states --represented by points of $\Gamma \subset\Gammaq$-- whose full quantum dynamics is extremely well-approximated by the quantum corrected trajectories on $\Gammac$. 

The key question is whether one can find the desired embedding of $\Gammac$ into $\Gammaq$ by locating its appropriate sub-manifold $\Gamma$.%
\footnote{This task is somewhat analogous to that of finding the `trial wave function' in the variational method in the standard perturbation theory of quantum mechanics. It is a mixture of science and art!}
An ideal example is a harmonic oscillator (or a linear field theory in a
stationary space-time). In this case, $\Gamma$ can be taken to be the
sub-manifold of quantum states consisting of the standard coherent states, which
is naturally isomorphic to $\Gammac$ because each coherent state is completely
determined by the point of $\Gammac$ at which it is peaked. As is well-known,
the full quantum evolution generated by $\Xq$ is in fact exactly tangential to this $\Gamma$. In this case, the quantum corrected dynamical trajectories capture \emph{full} quantum dynamics. In simple cosmological models, one can find an embedding of $\Gammac$ to $\Gammaq$ such that $\Xq$ is tangential to $\Gamma$ to a high degree of accuracy (which can be specified precisely \cite{jw,vt}). This $\Gamma$ is also spanned by a family of carefully chosen, sharply peaked Gaussian quantum states $\Psi$ in $\Gammaq$. The resulting quantum corrected dynamical trajectories in $\Gammac$ approximate the full quantum evolution of these $\Psi$ to a high degree of accuracy even in the Planck regime \cite{vt,asrev}. These trajectories again define certain quantum corrected FLRW metrics, denoted $\gb_{ab}$, whose coefficients now depend on $\hbar$.  The evolution of the scale factor of $\gb_{ab}$ becomes significantly different from that in classical general relativity once the expectation value of matter density (or curvature) enters the Planck regime, leading to the resolution of the big-bang singularity. 

By construction, the phase space of the effective theory is the same as in the classical theory, but the effective Hamiltonian constraint (generating the projection of the Hamiltonian flow $\Xq$ into $\Gamma$) is different. When tailored to the proper time $t$ of the FLRW space-times, it is given by \cite{vt,acs,asrev}:
\be
\label{ehc} \heff := \f{p_\phi^2}{4\pi G v}\, -\, \f{3 v}{4 \l^2}\, \sin^2(\l h)\approx0. \ee
The leading order quantum corrections are encoded in (\ref{ehc}) through $\l$; in the limit $\l \to 0$, one recovers the classical Hamiltonian constraint (\ref{chc}). To obtain the \EE governing the space-time geometry, one first obtains the equations of motion starting from the effective Hamiltonian (\ref{ehc}):
\ba \label{effeqs}
\dot v &=& \{v,\,\heff\}=\f{3 v}{2 \l} \sin(2 \l h)\,, \qquad 
\dot\phi = \{\phi,\, \heff\} = \f{p_{\phi}}{2\pi G v}\, ,
\nonumber \\
\dot{h} &=& \{h,\,\heff\}= \f{3}{2}\, \f{\sin^{2}(\l h)}{\l^{2}} + \f{p_\phi^2}{2\pi G v^2}\,, \qquad \dot{p}_{\phi} = \{p_{\phi},\, \heff\} = 0\, .
\ea
Let us set 
\be \rho := \f{p_{\phi}^{2}}{2V^{2}} \qquad {\rm and} \qquad \rhosup  := \f{3}{8\pi G \l^{2}} \equiv \f{18\pi}{G^{2}\hbar\,\Delta_{o}^{3}} \, ,\ee 
so that $\rhosup$ is the upper bound of the density operator $\h\rho$ on the
physical Hilbert space \cite{acs}. For concreteness, one generally uses the
value $\Delta_{o} = 5.17$ by appealing to black hole entropy calculation in LQG, which yields $\rhosup \approx 0.41 \rho_{\rm Pl}$. 

Using (\ref{ehc}) and (\ref{effeqs}) we obtain the following quantum corrected or effective Friedmann equation:
\be
\label{fried} H^2 = \Big(\f{\dot v}{3 v}\Big)^2 = \f{8\pi G}{3}
\rho\Big(1-\f{\rho}{\rhosup}\Big)\, ,
\ee
where $H$ is the Huuble rate. Similarly, by taking the time derivative of (\ref{fried}) and using the equation of motion for $\phi$ we obtain the quantum corrected Raychaudhuri equation%
\be
\label{ray}  \dot H = -8 \pi G \rho\, \Big(1- 2\f{\rho}{\rhosup}\Big)\, \equiv\, -4 \, \pi G\, (\rho\,+p)\, \Big(1- 2\f{\rho}{\rhosup}\Big)\, ,
\ee
where in the last step we cast the equation in the more familiar form using the
fact that $p=\rho$ for massless scalar fields, $p$ being the pressure of the
scalar field. Eqs (\ref{fried}) and (\ref{ray}) are the \EE that govern the quantum corrected geometry corresponding to sharply peaked Gaussian states $\Psi$ in $\Hp$ of LQC. Note that $\rhosup$, the upper bound of the density operator $\h{\rho}$ in the full quantum theory,  plays a direct role in the effective dynamics. In particular, since the Hubble parameter vanishes at $\rho=\rhosup$, the effective dynamical trajectories bounce there. Finally, since (\ref{fried}) and (\ref{ray})  do not refer to the fiducial cell at all, we can trivially remove the infrared cut-off by letting the cell occupy all of $\mathbb{R}^{3}$.

Note that as the area gap goes to zero, $\rhosup$ diverges and we recover the standard Friedmann and Raychaudhuri equations of general relativity. These \EE extend to more general models with appropriate modifications \cite{awe2,madrid-bianchi,awe3,we}. They have been extremely useful in providing physical intuition for the mechanism underlying singularity resolution because they involve $\hbar$-dependent but \emph{smooth space-time fields}, rather than the more abstract wave functions $\Psi$. These equations have often provided the first glimpses of the novel aspects of Planck scale physics contained in the sharply peaked states $\Psi$, which was later confirmed by numerical simulations of full quantum equations.

\section{Generalized effective equations}
\label{s3}
 
Recently, numerical simulations were carried out to probe the quantum dynamics of states $\Psi$ which are \emph{not} sharply peaked \cite{dgs1,dgms}. Specifically, one considers states $\Psi$ --both Gaussian and non-Gaussian-- which have large dispersions in the volume operator $\h{V}\mid_{\phi_{o}}$ at a late time $\phi_{o}$, evolves them back in time, and evaluates the expectation values of the principal Dirac observables $\h{p}_{\phi},\, \h{V}\mid_{\phi}$ for earlier times $\phi$. These evolutions became feasible thanks to the introduction of new numerical techniques in a framework called the ``Chimera numerical scheme'' \cite{dgs1}. As mentioned in  section \ref{s1}, these simulations showed that while the dynamics of the expectation values of $\h{p}_{\phi}$ and $\h{V}|_{\phi}$ is qualitatively similar to that in sharply peaked states which were studied earlier \cite{aps3}, there are major departures in the detailed quantitative behavior. In particular, the bounce continues to occur for states with wide spreads --as is obvious from Eq (\ref{V}). But the details of dynamics are very different from those predicted by \EE (\ref{fried}) and (\ref{ray}). In particular, the  matter density at the bounce can be significantly lower and there is a  clear pattern: more the state is spread at late times, smaller the density at the bounce. These numerical results naturally led to a number of questions. Are there \emph{generalized effective equations} that correctly capture the quantum dynamics of these expectation values? Put differently, can one derive the equations governing these \emph{mean-value trajectories} (MVTs) from first principles? Do these MVTs also define a smooth FLRW geometries in a consistent fashion? Are there generalizations of (\ref{fried}) and (\ref{ray}) that capture the key quantum corrections induced by the widely spread states $\Psi(h,\phi)$? Since $\rhosup$ is the upper bound of the density operator $\h{\rho}$ on the physical Hilbert space $\Hp$, it is not surprising that for general widely spread states the density at the bounce, $\rho_{B}$ is less than $\rhosup$. But is there a formula that quantitatively expresses the anti-correlation between the spread of the state $\Psi$ and the density at the bounce? We will answer these questions in this and the next section.

Let us then consider general quantum states $\Psi(h,\phi)$ in the physical Hilbert space $\Hp$. The expectation values of the Dirac observables $\h{V}|_{\phi}, \h{p}_{\phi}$ have a well-defined evolution with respect to the internal time $\phi$. We will denote them by $\Vb$ and $\b{p}_{\phi}$. Thus, given any physical state $\Psi$, Eq. (\ref{V}) provides us a MVT given by:
\be \label{meanevo} \Vb(\phi) = V_{+} e^{\alpha \phi} + V_{-} e^{-\alpha \phi}, \qquad \b{p}_{\phi} (\phi) = \b{p}^{o}_{\phi} \ee
where $\alpha=\sqrt{12\pi G}$ as before, and $\bar{p}^{o}_{\phi}$ and $V_{\pm}$ are constants determined by $\Psi$ \cite{acs,asrev} (see Appendix A). The key questions are: i) Can one assign to each of these trajectories a smooth, quantum corrected FLRW metric $\b{g}_{ab}$ consistently, such that $\phi$ satisfies the wave equation $\bar{\Box} \phi =0$; and, if so, ii) What are the quantum corrected equations governing $\gb_{ab}$? We will show that the answer to the first question is in the affirmative, and the answer to the second question provides the generalization of the standard LQC \EE (\ref{fried}) and (\ref{ray}) that hold for all states $\Psi$. 

To specify the desired \emph{mean-value FLRW metric} $\b{g}_{ab}$, we need to specify the relation between $\phi$ and the proper time $\b{t}$ and specify the time-dependent scale factor $\ab$ of $\gb_{ab}$. The intuition derived from the classical phase space formulation of general relativity summarized in section \ref{s2.1} suggests that we define $\tb$ and $\ab$ via:
\be \label{bar} \f{\rmd \phi}{\rmd \tb} := \f{\pb_{\phi}^{o}}{\Vb (\phi)},
\qquad {\rm and} \qquad \ab (\phi) := \Big(\f{\Vb(\phi)}{V_{o}}\Big)^{\f{1}{3}} \ee   
where $V_{o}$ is the co-moving volume of the cell $\mathcal{C}$. We can then consider a 4-manifold, topologically $\mathbb{R}^{4}$ and define on it the FLRW metric
\be \label{gb} \gb_{ab}\rmd x^{a} \rmd x^{b} \, =\, - \rmd \tb{}^{2} + \ab^{2}(\tb)\, \rmd \vec{x}^{2} \, . \ee
We will refer to $\gb_{ab}$ as the \emph{mean value FLRW metric} determined by the given quantum state $\Psi$. Is there a simple equation that $\phi$ automatically satisfies on this mean value FLRW space-time? Note first that (\ref{bar}) and (\ref{V}) immediately imply
\be \partial_{\tb}\phi\, =\, \pb_{\phi}^{o}\, \big[V_{+}\, e^{\alpha\phi} +
V_{-}e^{-\alpha \phi}\big]^{-1}\, . \ee
By taking the second time derivative, one finds that $\phi$ satisfies the massless Klein-Gordon equation with respect to the mean value metric $\b{g}_{ab}$:
\be \label{kg} \b{\Box} \phi = \partial_{\tb}^{2}\, \phi^{2} \, +3 \b{H}\, \partial_{\tb}\phi = 0, \ee
where $\b{H} := ({\partial_{\tb} \Vb}/{3\Vb})$ is the Hubble parameter of the mean-value metric.

Next, let us compute the equations that govern the dynamics of the mean value metric $\gb_{ab}$ itself. To obtain these equations it is natural to define the mean value matter density $\b\rho$ and pressure $\pb$ as  
\be \label{density} \b\rho = \f{1}{2}\, \f{(\pb_{\phi}^{o})^{2}}{\Vb^{2}}\qquad {\rm and} \qquad \pb = \b\rho\, , \ee
since $\phi$ satisfies the massless Klein Gordon equation with respect to $\gb_{ab}$. Using (\ref{V}) it is easy to verify that the expectation value of the volume bounces in any given state $\Psi$, and at the bounce the mean value volume and hence the mean value matter density are given by
\be \label{meanbounce} \Vb_{\rm B} = 2\sqrt{V_{+}V_{-}} \quad {\rm and} \quad \rhob=  \f{(\pb_{\phi}^{o})^{2}}{8 V_{+}V_{-}}\, . \ee
With these notions at hand, we can obtain the generalized effective Friedmann equation using (\ref{bar}): 
\ba \label{genfried} \b{H}^{2} &:=& \Big(\f{\partial_{\tb}\Vb}{3\Vb}\Big)^{2}\, =\, \f{8\pi G\b\rho}{3}\, \Big(1\, -\, \f{4V_{+}V_{-}}{\Vb^{2}}\Big) \nonumber\\
&=&  \f{8\pi G\b\rho}{3}\, \Big(1\, -\, \f{\b\rho}{\rhob}\Big)\, .\ea

By taking the time derivative of $H$ and using (\ref{V}), one obtains the generalized effective Raychaudhuri equation. Using the fact that $\pb = \b\rho$, it can be cast in the form
\be
\label{genray} \dot H = -4\pi G (\b\rho + \pb) \left(1 - 2 \f{\b\rho}{\rhob}\right),
\ee
or, alternatively, as an equation involving the second time derivative of the mean value scale $\ab$:
\be 
\f{\ddot \ab}{\ab} = - \f{4\pi G}{3} \b\rho \left(1 - 4 \f{\b\rho}{\rhob}\right)
- 4\pi G \pb \left(1 -2 \f{\b\rho}{ \rhob}\right),
\ee
which brings out the fact that it reduces to the familiar Raychaudhuri equation
of general relativity as $\rhob \to \infty$. Thus, in the mean value
approximation, the dynamics encapsulated in a general quantum state $\Psi$ is
specified by the coupled, non-linear system of equations (\ref{kg}),
(\ref{genfried}) and (\ref{genray}). The form of these equations is similar to
the system consisting of the Klein-Gordon, Friedmann and Raychaudhuri equations of classical relativity but there are $\hbar$-dependent quantum corrections, encoded in $\rhob$, that carry the signature of the quantum state $\Psi$. A non-trivial feature of these equations is that they are not all independent, but constitute a consistent set: For example, as in classical general relativity, the (generalized, effective) Raychaudhuri equation can be derived from the (generalized, effective) Friedmann equation and the Klein-Gordon equation. Finally, since $\rhosup$ is the upper bound on the spectrum of the density operator $\h{\rho}$ on the physical Hilbert space, it follows that $\rhob \le \rhosup$. Comparing the \GEE with the \EE in section \ref{s2.2}, it then follows that the quantum gravity corrections set in \emph{earlier} --i.e., at lower densities and curvatures-- for the widely spread quantum states than they do for the sharply peaked ones. We will return to the last two issues in section \ref{s4}.

Let us summarize. We began with a general quantum state $\Psi$ in the physical Hilbert space. We know from \cite{acs} that the mean values, $\Vb,\,\pb_{\phi}$ of the principal Dirac observables evolve via (\ref{meanevo}). We showed that this MVT in phase space variables $(\b{v}, \pb_{\phi}, \phi)$ naturally defines a mean value FLRW metric $\gb_{ab}$ given in (\ref{gb}) and (\ref{bar}) that satisfies the GEE (\ref{genfried}) and (\ref{genray}). Furthermore, the field $\phi$ we began with automatically satisfies the Klein-Gordon equation with respect to this $\b{g}_{ab}$. Finally, since the \GEE make no reference to the fiducial cell, they remain unaltered in the infinite volume limit.
In this precise sense, the quantum state $\Psi$ leads to a self-consistent mean value space-time description of gravity coupled to a Klein Gordon field. It captures the dynamics of expectation values of the principal Dirac observables \emph{exactly} without any approximations. Interestingly, the information about the quantum state $\Psi$ is encoded in these \GEE only through the mean value $\rhob$ of the matter density at the bounce. It is rather striking that the \GEE differ from the \EE (\ref{fried}) and (\ref{ray}) only through the replacement of the constant $\rhosup$ by the $\Psi$-dependent parameter $\rhob$. Note however that the \GEE have no information at all about dynamics of the fluctuations of these Dirac observables. For widely spread states these fluctuations are large and therefore the mean value space-time description misses a great deal of information about full quantum dynamics of $\Psi$. Nonetheless, the \GEE are useful for at least two reasons. First, they provide a rather simple analytical understanding of the results of numerical simulations of Refs \cite{dgs1,dgms}. Second, as discussed in section \ref{s4.3}, these equations also facilitate the analysis of the pre-inflationary dynamics of cosmological perturbations.

\section{Features of the Generalized Effective Equations}
\label{s4}

This section is divided into three parts. In the first we obtain the Hamiltonian constraint governing the \GEE by exploiting an interplay between the area gap and the value of the matter density at the bounce. In the second we discuss the relation between the matter density at the bounce and the dispersions in the volume at late time. In the third, we report results that show that, even when the dispersions in volume are significant at late times, the mean value metric $\gb_{ab}$ can provide a good approximation to the `dressed, quantum corrected metric' used in the analysis of pre-inflationary dynamics of cosmological perturbations. 

\subsection{The area gap and the generalized effective Hamiltonian}
\label{s4.1}

Recall from section \ref{s2.2} that the interplay between the quantum phase space $\Gammaq$ and the classical phase space $\Gammac$ provides an effective Hamiltonian constraint (\ref{ehc}) which then leads to the standard \EE (\ref{fried}) and (\ref{ray}). However, this interplay depends on the availability of a suitable embedding of the classical phase space $\Gammac$ into the quantum phase space $\Gammaq$,  and the image $\Gamma \subset \Gammaq$ of $\Gammac$ is spanned by sharply peaked states $\Psi$. That is why, while the \EE\, (\ref{fried}) and (\ref{ray}) provide an excellent approximation to full quantum dynamics, they do so only for these very special $\Psi$. The \GEE (\ref{genfried}) and (\ref{genray}), on the other hand, were derived through an entirely different strategy. We allowed general states $\Psi$ and used only the equations that govern the dynamics of the \emph{expectation values} of the principal Dirac observables $\h{p}_{\phi}$ and $\h{V}|_{\phi}$. Nonetheless given the close similarity in the forms of \GEE and the standard EEs, it is natural to ask if the \GEE can also be derived from a generalized effective Hamiltonian. We will now show that this is indeed possible, thanks to an interesting interplay between the density at the bounce and the area gap.

Let us begin by noting that the leading quantum corrections are encoded in the \EE\,  (\ref{fried}) and (\ref{ray}) via $\rhosup = 18\pi/(G^{2}\hbar \Delta_{o}^{3})$. Furthermore, the only difference in the forms of the \EE and the \GEE is the replacement of $\rhosup$ by $\rhob$. Finally neither set of equation features the area gap $\ag$ (or $\l$) explicitly. These three features motivate the introduction of a fictitious LQC theory in which the area gap $\ag$ is replaced by $\Delta$: 
\be \label{Delta} \Delta\, =\, \Big(\f{18\pi}{G^{2}\hbar\rhob}\Big)^{\f{1}{3}}\,\, \equiv \,\,
\Big(\f{18\pi\rhopl}{\rhob}\Big)^{\f{1}{3}} \ee
Then, we can use geometric quantum mechanics in this fictitious $\Delta$-LQC and obtain the \EE following the procedure 
that was used to obtain the standard \EE in the $\ag$-LQC \cite{vt,asrev}. This purely mathematical strategy will then lead us precisely to our \GEE equations (\ref{genfried}) and (\ref{genray})  of the $\ag$-LQC. Therefore, it follows that our \GEE of section \ref{s3} are indeed governed by a generalized effective Hamiltonian constraint
\be 
\label{gehc} \heffgen = \f{\pb_\phi^2}{4\pi G|\bar{v}|}\, -\, \f{3 |\bar{v}|}{4 \ell^2}\, \sin^2(\ell \bar{h})\, \approx 0\,, \quad {\rm where} \quad \ell^{2} = \f{\Delta^{3}}{48\pi^{2}}\, \lp^{2} \ee
Thus, the only difference between $\heff$ and $\heffgen$ is that $\l$ in (\ref{ehc}) is replaced by $\ell$ in (\ref{gehc}). As in section \ref{s2.2}, we can now obtain equations of motion for phase space variables, now denoted by $\bar{v}, \bar{h}; \phi,  \bar{p}_{\phi}$, and derive from them the  equations governing $\phi$ and the mean value metric $\gb_{ab}$. These are precisely the Klein-Gordon equation (\ref{kg}) for $\phi$ and the \GEE (\ref{genfried}) and (\ref{genray}) for $\gb_{ab}$. 

To summarize, the mean value metric $\gb_{ab}$ defined by a general quantum state $\Psi$ in the standard $\ag$-LQC obeys \GEE which are identical to the standard \EE in a fictitious $\Delta$-LQC, where $\Delta$ is determined by the given $\Psi$ via (\ref{Delta}). Note, however, that the \EE in the $\Delta$-theory accurately capture the full quantum dynamics only for those states which are sharply peaked in the $\Delta$-theory. They do not inform us about the quantum evolution of the given widely spread state $\Psi$ beyond the expectation values of the principal Dirac observables $\h{p}_{\phi}$ and $\h{V}|_{\phi}$; they have no information about the dynamics of `higher moments' of $\Psi$ such as the uncertainties $\big(\langle \h{V}^{2} \rangle_{\phi} - \langle \h{V}\rangle^{2}_{\phi} \big)$.

\subsection{Relation between $\rhob$ and $(\Delta V)/\Vb$}
\label{s4.2}

As noted above, the only difference between the \EE (\ref{fried}) and (\ref{ray}) and the \GEE (\ref{genfried}) and (\ref{genray}) is that $\rhosup$ is replaced by $\rhob$. The standard \EE govern the dynamics of states $\Psi_{\rm peak}$ that are sharply peaked in volume at late times, while the \GEE hold for more general states $\Psi_{\rm gen}$ which can have arbitrary spread in volume at late times. Therefore, one would expect the standard \EE to be a special case of the \GEE. Indeed, at an intuitive level, it is clear that this is the case because numerical simulations have shown that the sharply peaked states bounce when $\rho = \rhosup$ \cite{aps3}. However, as we explained in section \ref{s4.1}, the two sets of equations are derived via entirely different avenues and their physical meanings are also rather different. Physically, $\Psi_{\rm peak}$ remain sharply peaked in both the Dirac observables and the standard \EE track their peaks. $\Psi_{\rm gen}$ on the other hand may have no peak or multiple peaks and the \GEE track only the evolution of the expectation values of $\h{p}_{\phi}$ and $\h{V}|_{\phi}$. Therefore, to establish a clear relation between the two sets of equations, we will now restrict ourselves to states $\Psi_{\rm gen}$ that are Gaussian (in the precise sense spelled out below) but can have large dispersions in volume at late times. For this class, we will be able to establish a relation between the matter density at the bounce with the dispersion in the volume at late times. It will follow, in particular, that as the dispersion is reduced, the density $\rhob$ at the bounce increases and tends to $\rhosup$.

Let us denote the expectation value of the Dirac observable $\h{p}_{\phi}$ by\, $\pb^{o}_{\phi} =: \hbar k_{o} (1+\sigma^2/4k_o^2)$\, and the dispersion in $p_{\phi}$ by\, $(\Delta p_{\phi})=(\hbar\sigma/2)\sqrt{1-\sigma^2/4k_o^2}$\, \cite{Corichi:2011rt} (also see Appendix A). Because $\h{p}_{\phi}$ is a constant of motion, $k_{o}$ and $\sigma$ are time independent and they determine the Gaussian state completely. We will restrict ourselves to physical states $\Psi$  for which $\sigma \ll k_{o}$ and $\alpha\equiv \sqrt{12\pi G} \ll k_{o}$. (Both $k_{o}$ and $\sigma$ have the same physical dimensions as $\alpha$.) Thus, all our states in this sub-section will be sharply peaked in $\h{p}_{\phi}$. However, there is no restriction on the dispersions in volume (or matter density). This is possible because  $\h{p}_{\phi}$ and $\h{V}|_{\phi}$ fail to commute on physical states.

As shown in Appendix A, at late times the relative dispersion in the volume in this state is given by \cite{Corichi:2011rt}
\be \label{DeltaV}
  \left(\f{\Delta V}{\Vb}\right)^2_{\rm late\,\,\, time}\!\!\!\!\! =\,\,\,\, e^{\alpha^2/\sigma^2}\,\, 
  {\left(1+\f{3\sigma^2+4\alpha^2}{4k_0^2}\right)}
              \left(1+\f{\sigma^2+\alpha^2}{4k_0^2}\right)^{-2} - 1 \,\,\,\,\approx\,\,\,\, e^{\alpha^2/\sigma^2} - 1\, ,  
\ee
where the last approximate equality follows from the fact that the round brackets in the first step are approximately equal to 1  
by our assumption on $\sigma/k_{o}$ and $\alpha/k_{o}$. Thus, the late time
uncertainty in volume is dictated by the relative magnitude of $\alpha$ and
$\sigma$. The state $\Psi$ is sharply peaked in volume at late times if $\alpha
\ll \sigma$, and is widely spread if $\alpha > \sigma$. For example, the early
simulations of this model \cite{aps2,aps3} were carried out using $\sigma/k_{o}
= 5\times 10^{-2}$ and  $\alpha/\sigma = 2.5\times 10^{-2}$. For these values we have the following relative dispersions:
\be 
\label{dvv1}\f{\Delta p_{\phi}}{p_{\phi}} \approx 2.5 \times 10^{-2}\quad {\rm
and} \quad \Big( \f{\Delta V}{\Vb}\Big)_{\rm late \,\,\, time} \approx
3.5 \times 10^{-2}. \ee
\begin{figure}[]
  \begin{center}
  \vskip-0.4cm
    \includegraphics[width=0.45\textwidth]{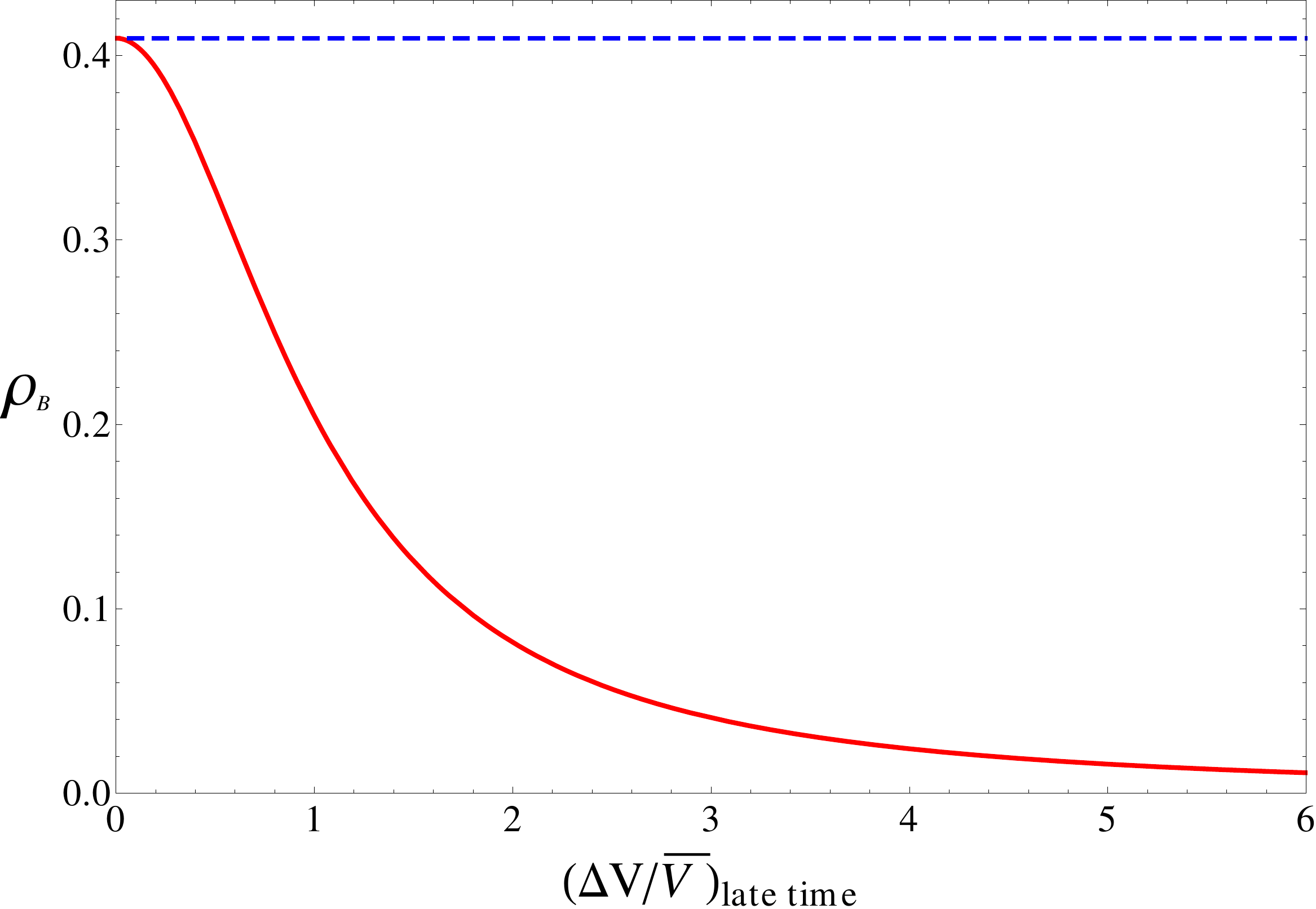}
    \includegraphics[width=0.45\textwidth]{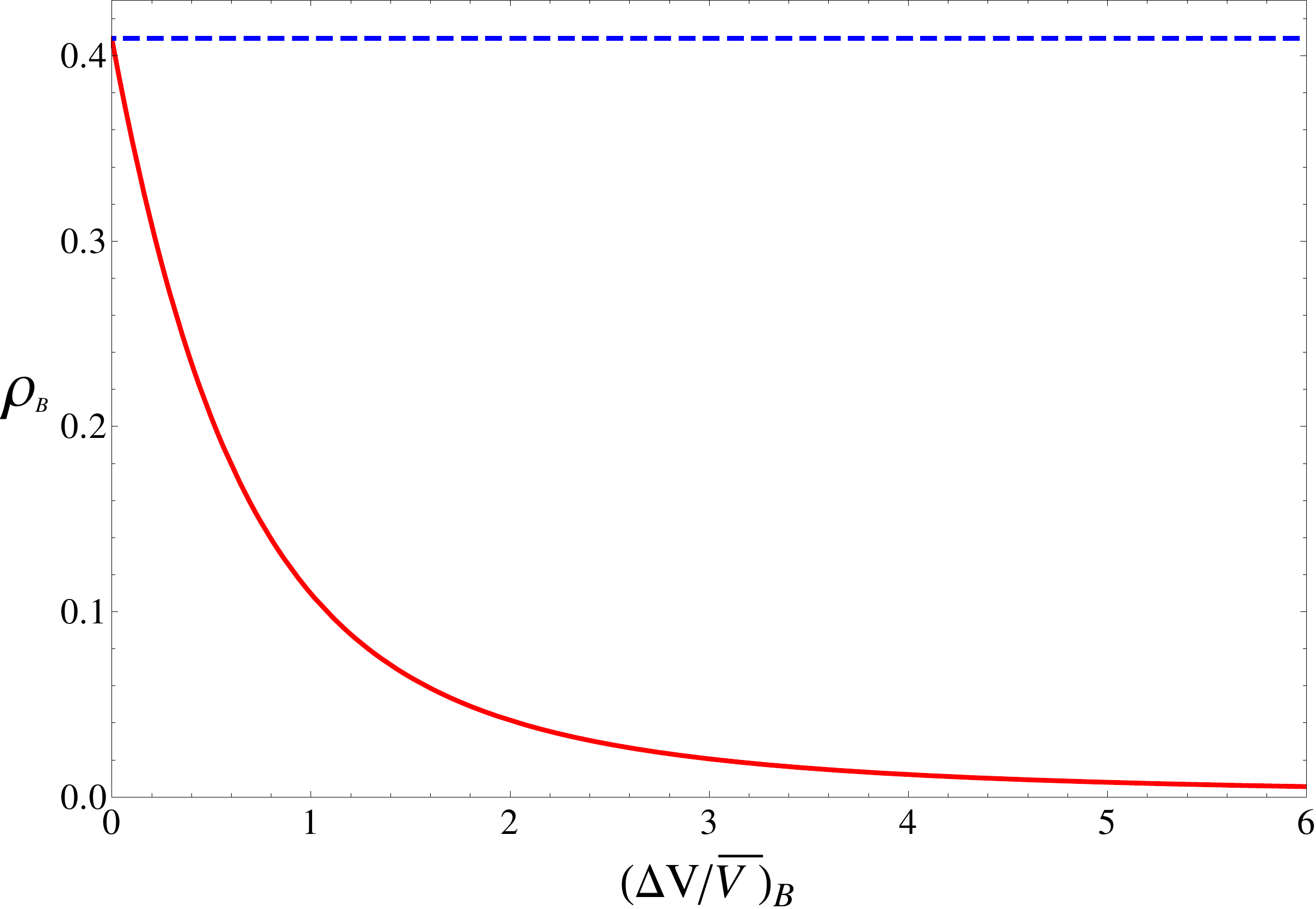}
\caption{\textit{Left Panel:} The plot of the mean value density $\b\rho_{\rm B}$ at the bounce against the relative dispersions in volume \emph{at late times} in a family of Gaussian states. Here, the parameter $k_{o}$ of the Gaussian states was set to $k_o=1000$ (in Planck unit), and the parameter $\sigma$ varies with $(\Delta V/\overline V)$. {\textit{Right Panel:}} The plot of the mean value density $\b\rho_{\rm B}$ at the bounce against the relative dispersions in volume \emph{at the bounce} for the same family of Gaussian states as was used in the left panel.}
\label{1}
\end{center}
\end{figure}
\noindent More recent simulations \cite{dgs2,dgms} considered $\sigma/k_{o} =
1.2 \times 10^{-3}$ and $\alpha/\sigma =1.02$, leading to states in which there is a wide spread in the relative volume at late times:
\be 
\label{dvv2}\f{\Delta p_{\phi}}{p_{\phi}} \approx 3 \times 10^{-3}\quad {\rm and} \quad \Big( \f{\Delta V}{\Vb}\Big)_{\rm late \,\,\, time} \approx 1.35. \ee
Therefore, these states do not qualify as semi-classical. Next, let us examine the density at the bounce. From \cite{Corichi:2011rt} (also Appendix A), we have
\be \b\rho_{\rm B}\,\, :=\,\, \f{(\pb^{o}_{\phi})^{2}}{8V_{+}V_{-}}\,\, =\,\, \rhosup\,\,\, e^{-\f{\alpha^{2}}{\sigma^{2}}}\,\, 
  {\left(1+\f{\sigma^2}{4k_0^2}\right)^{2}}
              \left(1+\f{\sigma^2+\alpha^2}{4k_0^2}\right)^{-2}  \, .
\ee
Hence, using (\ref{DeltaV}) we conclude that 
\ba \label{rel} \f{\b\rho_{\rm B}}{\rhosup}\, &=&\, {\left(1+\f{\sigma^2}{4k_0^2}\right)^{2}\left(1+\f{3\sigma^2+4\alpha^2}{4k_0^2}\right)}{\left(1+\f{\sigma^2+\alpha^2}{4k_0^2}\right)^{-4}}\,\, \left(1+ \Big(\f{\Delta V}{\Vb}\Big)^2_{\rm late\,\,\, time}\right)^{-1} \,\,\, \nonumber\\
&\approx&\,\,\, \left(1+\Big(\f{\Delta V}{\Vb}\Big)^2_{\rm late\,\,\, time}\right)^{-1}\, ,
\ea
where in the last approximate equality we have again used our assumptions on $\sigma/k_{o}$ and $\alpha/k_{o}$. In particular, for two examples we considered, we have the following results: for the sharply peaked state considered in \cite{aps2,aps3}, we have $\b{\rhob} \approx 0.99 \rhosup$ and for the state with wide spread considered in \cite{dgs2,dgms}, we have $\b\rho_{\rm B} \approx 0.35 \rhosup$. Eq. (\ref{rel}) provides a precise relation between the spread in volume at late times and the density at the bounce for the general class of Gaussian states specified above. It brings out the fact that as we consider states that are more and more sharply peaked in volume at late times, the density $\b\rho_{\rm B}$ at the bounce tends to $\rhosup$, the upper bound in the physical Hilbert space of the density operator $\h\rho$. Note that one can easily construct states that are more and more sharply peaked in both $\h{p}_{\phi}$ and $\h{V}|_{\phi}$ by increasing $k_{o}$ such that $\alpha \ll \sigma \ll k_{o}$. Even in the example of sharply peaked states we considered from the early simulations \cite{aps2,aps3}, to maintain numerical accuracy, $k_{o}$ was chosen to be only $5\times 10^{3}$ in Planck units. If one were to model the universe more realistically, the relative fluctuations in the two Dirac observables would become astronomically small. Then, for the Gaussian states, the energy density $\b\rho_{\rm B}$ at the bounce would be indistinguishable from $\rhosup$.

\begin{figure}[]
  \begin{center}
    \includegraphics[width=0.45\textwidth]{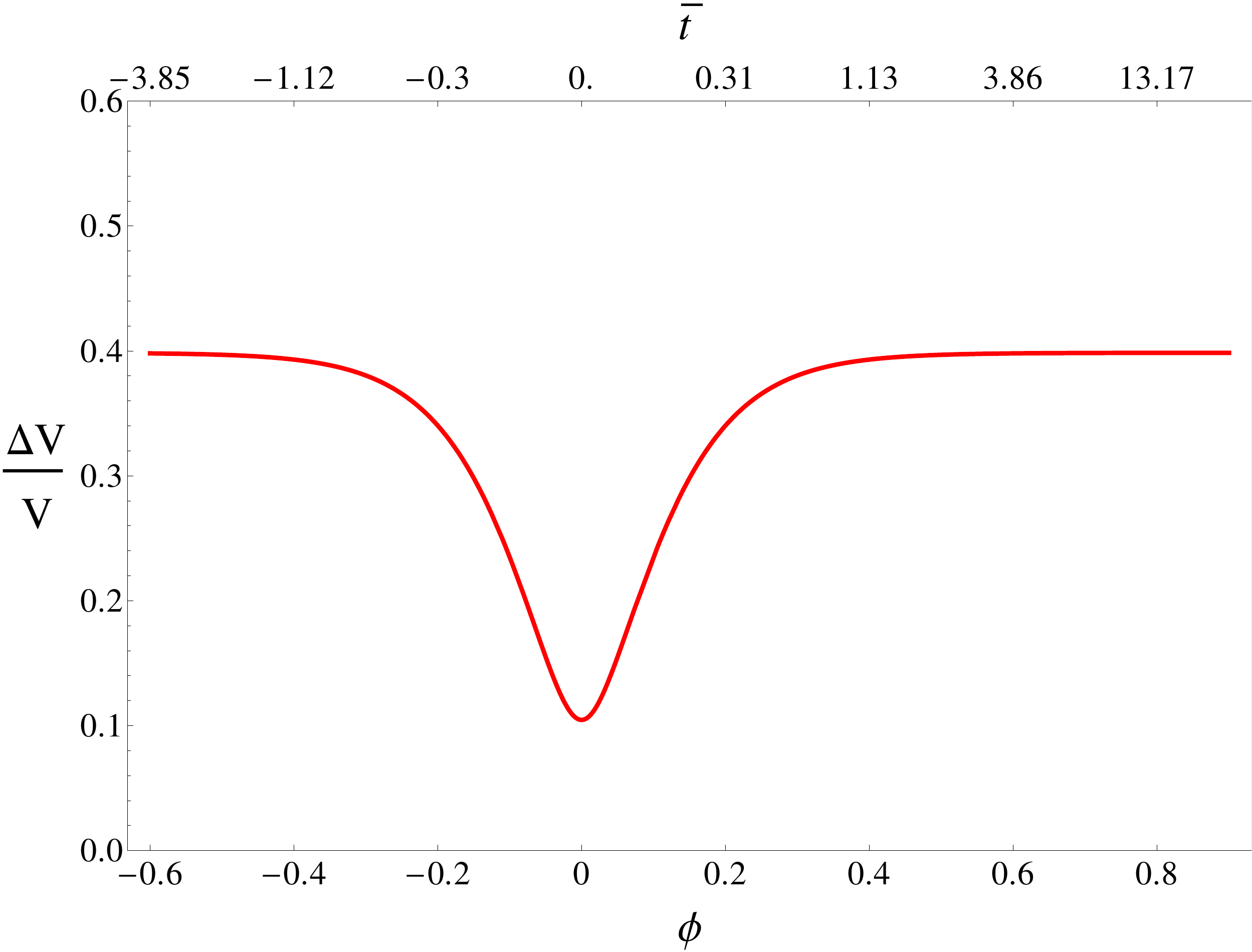}
    \hskip0.1cm
    \includegraphics[width=0.45\textwidth]{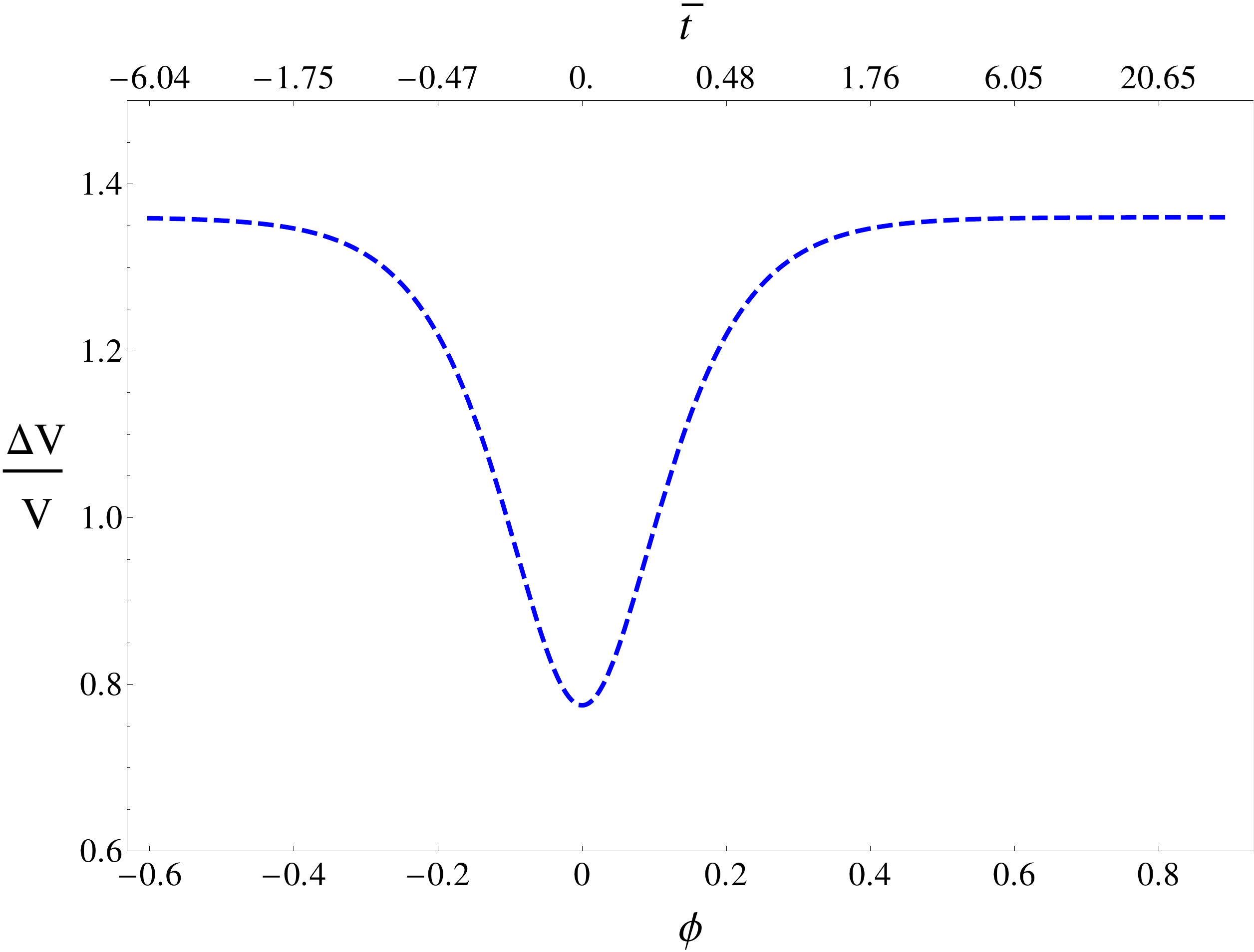}
\caption{ \textit{Left Panel:} The evolution of the relative dispersion $(\Delta
V/{\Vb})$ in volume for a Gaussian state with parameters: $k_o=1000$,
$\sigma=16$ (in Planck units)  and 
{\textit{Right Panel:}} parameters: $k_o=1000$, $\sigma=6$ (in Planck units).
The evolution is shown with respect to both the proper time $\bar t$ and the internal time $\phi$. The two plots bring out the fact that the asymptotic value of the relative dispersion is reached rather quickly for this family of states.}
\label{2}
\end{center}
\end{figure}
The left panel of Fig.~\ref{1} shows the exact relation --without the
approximation made in the second step in (\ref{rel})-- between the density $\b\rho_{\rm B}$ at the bounce and the relative dispersion in volume at late times, using a family of Gaussian states. One can see that $\b\rho_{\rm B}$ first decreases rapidly with $(\Delta V/{\Vb})$ and can become arbitrarily small for states which are extremely quantum mechanical just as one would expect from the $(1+(\Delta V/{\Vb})^2)^{-1}$ behavior. This behavior highlights the fact that the departures from classical general relativity can occur in tame regimes if the states are highly non-classical. In this respect the situation is qualitatively the same as in ordinary quantum mechanics which admits states in which macroscopic objects have a significant probability of being in two widely separated places at the same time. It is just that such states are not seen in Nature. Similarly, the cosmological model under consideration also admits states $\Psi$ in which the universe can bounce at the density of water. But such states are so extremely quantum mechanical that the relative uncertainty $(\Delta V/{\Vb})$ in volume in them would be $\sim 10^{47}$ at late times!  

In this discussion we considered the relative spread in volume at late times rather than at the bounce itself because the analog of (\ref{rel}) has a more complicated form in terms of $(\Delta V/{\Vb})_{\rm B}$. However, as the right panel of Fig. \ref{1} shows, the qualitative behavior is the same. Also, it is worth noting that the dispersion reaches its asymptotic values rather quickly in the Gaussian states considered here. This is illustrated in Fig.~\ref{2} which shows the evolution of the relative dispersion with the internal time $\phi$ as well as the proper-time $t$ for two different Gaussian states. Finally, for analytic control, we restricted ourselves to Gaussian states. But numerical simulations have been carried out also for non-Gaussian states with wide dispersions \cite{dgs2,dgms}. The qualitative behavior of $\b\rho_{\rm B}$ as a function of the relative dispersion in volume is the same. But, not surprisingly, our analysis of Gaussian states does not account for the detailed quantitative behavior.  

\subsection{Relation to the `dressed' metric}
\label{s4.3}

\begin{figure}
    \includegraphics[width=3.1in,height=2.3in,angle=0]{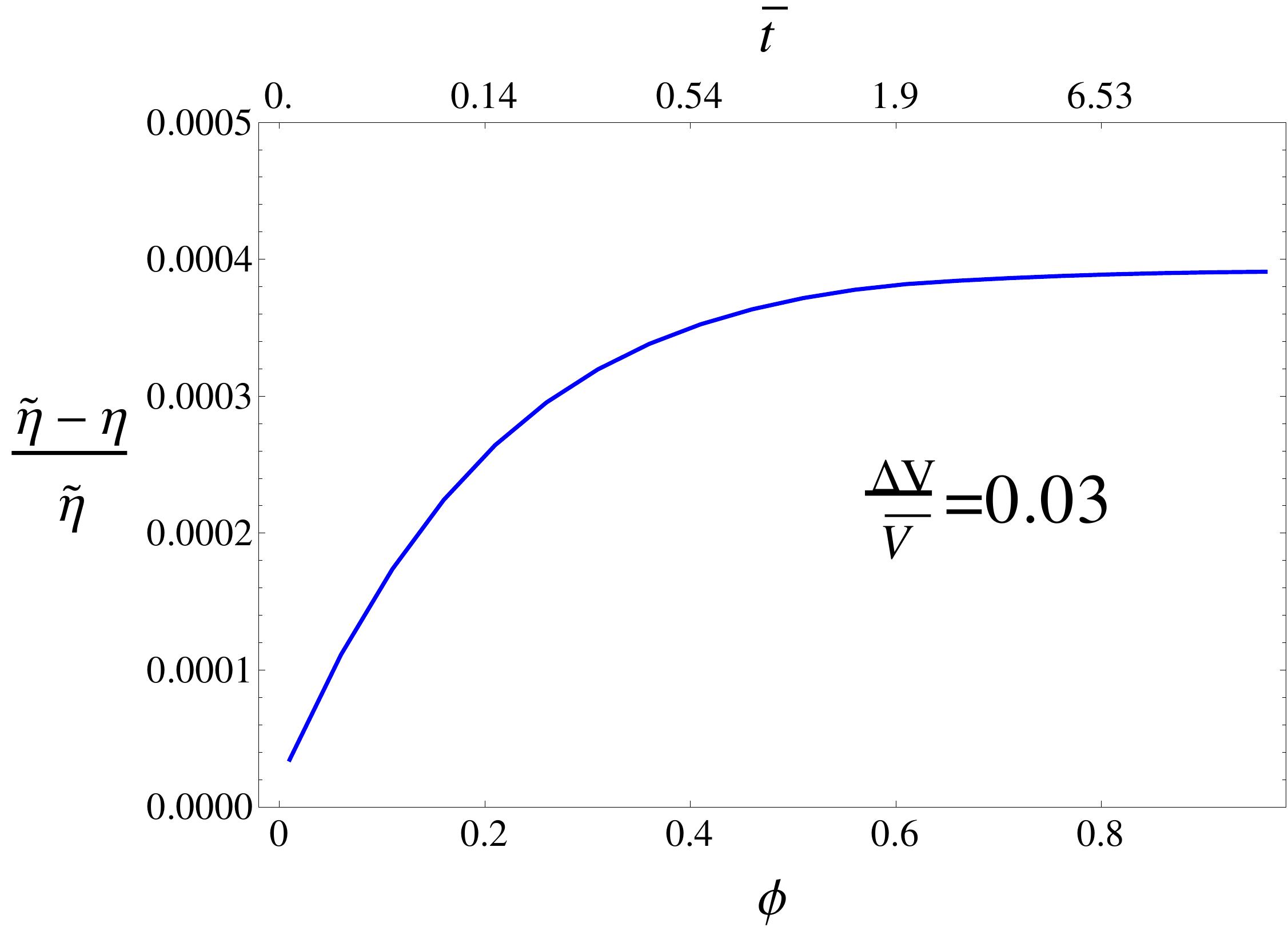}
    \includegraphics[width=3.1in,height=2.3in,angle=0]{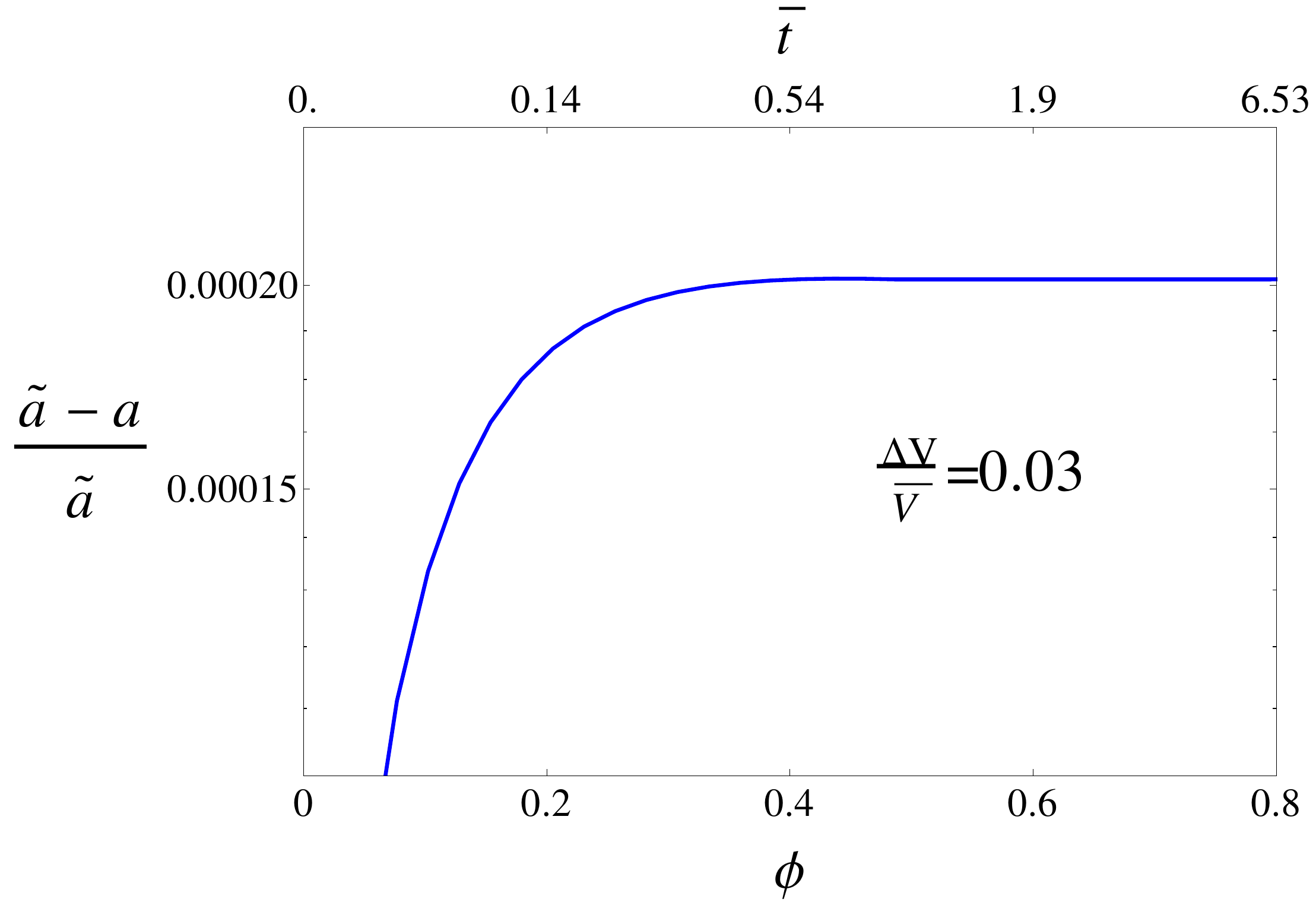}
\caption{Comparison of `dressed' and `mean value' quantities for a sharply peaked states with $\Delta V/\Vb=0.035$ considered in Eq. (\ref{dvv1}). The `internal' time is shown on lower x-axis, and the corresponding proper time $\bar t$ on the upper x-axis. {\it Left Panel:} Relative difference between the dressed and mean value conformal times $\tilde \eta$ and $\eta$. It is evident that the difference is less than $0.04\%$. {\it Right Panel:} Relative difference between the dressed and mean value scale factors $\tilde a$
and $a$. The difference is only $0.02\%$. This shows that for sharply peaked states considered in early simulations (such as the ones performed in \cite{aps2,aps3}) the effective metric is an excellent approximation to the dressed metric.}
\label{3}
\end{figure}
\begin{figure}[]
  \begin{center}
  \vskip-0.4cm
    \includegraphics[width=3.1in,height=2.3in,angle=0]{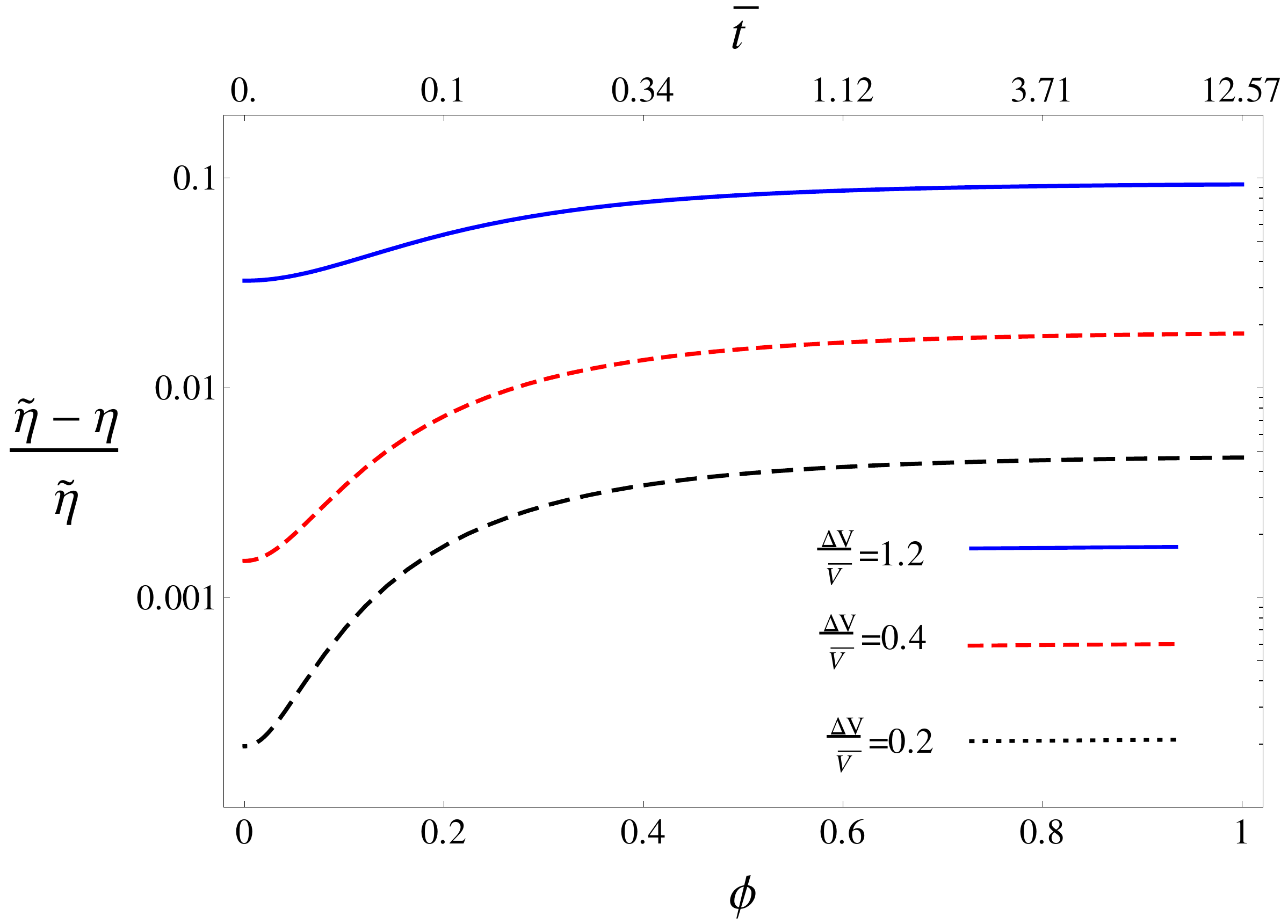}
    \includegraphics[width=3.1in,height=2.3in,angle=0]{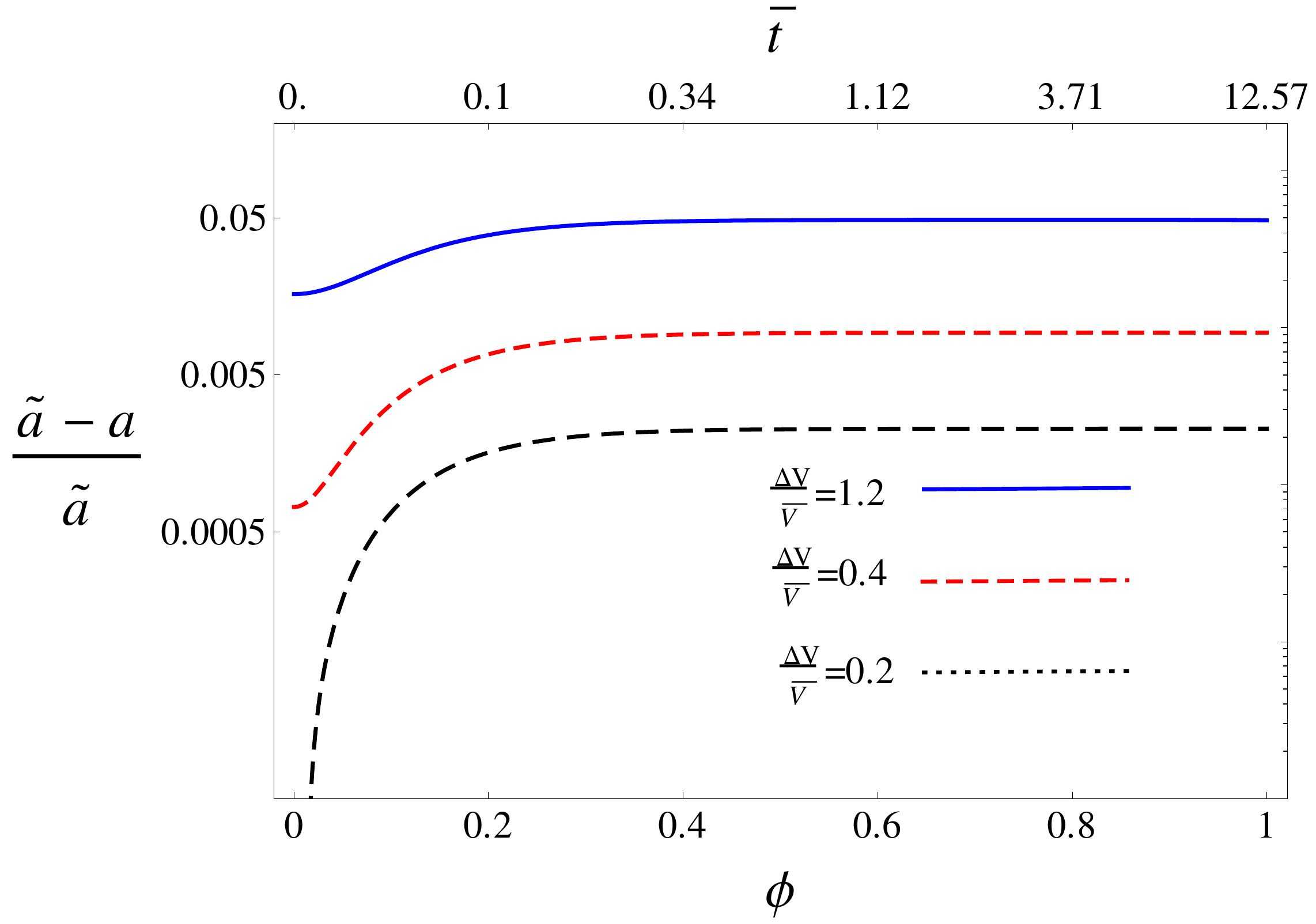}
\caption{ \textit{Left Panel:} The relative difference between the `dressed' and `mean value' conformal times $\t\eta$ and $\b\eta$ as a function of the `internal' time $\phi$ (the lower $x$ axis) and proper time  (the upper x-axis). The difference is only $\sim 0.5\%$ if the relative dispersion in volume at large times is $0.2$ but grows to $\sim 10\%$ as ispersion grows to 1.2. Here the parameters of the Gaussian states are: $k_o=1000/\alpha$, $\sigma=6.5/\alpha,\,15/\alpha$, and $30/\alpha$ (in Planck units) which correspond to the late time $\Delta V/\Vb$ values of $1.2,\,0.4$, and $0.2$ respectively.
{\textit{Right Panel:}} The relative difference between the `dressed' and `mean value' scale factors $\t{a}$ and $\b{a}$ as a function of the `internal' time $\phi$ (the lower x-axis) and proper time $\bar t$ (the upper x-axis). The difference is only $\sim 0.9\%$ if the relative dispersion in volume at late times is $0.2$ but grows to $\sim  5\%$ as dispersion grows to $1.2$. Both relative differences grow after the bounce but quickly reach their asymptotic values. For any given desired accuracy of results, these simulations determine the class of states $\Psi$ with wide relative dispersions for one can replace the `dressed' metric $\t{g}_{ab}$ with the mean value metric $\gb_{ab}$.}
\label{4}
\end{center}
\end{figure}

As mentioned in section \ref{s1}, because of singularity resolution, in LQC one can analyze the dynamics of the cosmological perturbations in the pre-inflationary epoch by facing the so-called ``trans-Planckian'' issues squarely \cite{aan1,aan2,aan3} (for a summary, see, e.g., \cite{abrev}). In the Planck regime of LQC, these perturbations are treated as quantum fields propagating on a \emph{quantum} FLRW geometry, described by a wave function $\Psi$. Detailed analysis has been performed both for the scalar modes $\R$ and the two tensor modes $T_{I}$, (with $I=1,2$). This analysis became possible because of an unforeseen simplification: It turns out that, so long as the back reaction on the wave function $\Psi$ of geometry can be neglected, dynamics of the quantum fields $(\h{\R}, \h{T}_{I})$ is \emph{completely} equivalent to that of quantum fields propagating on a smooth but \emph{quantum-corrected, `dressed' FLRW metric} $\t{g}_{ab}$\, determined by $\Psi$ \cite{akl}. The metric $\t{g}_{ab}$ arises from a detailed calculation and carries more information than the mean value metric $\b{g}_{ab}$; in particular, it `knows' about certain higher order fluctuations in the quantum FLRW geometry encoded in $\Psi$. Since it is customary to use conformal time in the theory of cosmological perturbations, 
$\t{g}_{ab}$ is expressed in terms of a `dressed' conformal time $\t\eta$ and a `dressed' scale factor $\t{a}$ as
\be \t{g}_{ab} {\rm d}x^a {\rm d}x^b = \t{a}^2 (-{\rm d}\t\eta^2 + {\rm d} \vec{x}^2)\, . \ee
In turn, $\t{\eta}$ and $\t{a}$ are constructed from the internal time $\phi$ and the $\phi$-dependent scale factor operator $\h{a} = (\h{V}/V_{o})^{1/3}$, where $V_{o}$ the co-volume of the fiducial cell $\mathcal{C}$. The precise relations are:   
\be \label{tilde} {\rm d}\bar{\eta} = \langle \h{H}_o^{-1/2}\rangle \, [\langle \h{H}_o^{-1/2}\h{a}^4 \h{H}_o^{-1/2}\rangle]^{1/2}\, {\rm d}\phi ; \quad{\rm and} \quad \bar{a}^4 = (\langle \h{H}_o^{-1/2}\, \h{a}^4 \h{H}_o^{-1/2}\rangle)/\langle \h{H}_o^{-1}\rangle \ee
where $\h{H}_{o} = \h{p}_{\phi}^{2}$ is the Hamiltonian governing the time evolution in relational time as in (\ref{hc1}). Thus, the dressed metric $\t{g}_{ab}$ is quite subtle to construct from the quantum state $\Psi$. Its form could not have been guessed; it arose from a detailed calculation showing the equivalence of the dynamics of the quantum fields $(\h{\R}, \h{T}_{I})$ on the quantum geometry defined by $\Psi$ and on the continuum FLRW geometry defined by $\t{g}_{ab}$

Because of the singularity resolution, space-time curvature remains finite in LQC, achieving its maximum value at the quantum bounce that replaces the big bang. This finiteness has interesting and unforeseen consequences on the dynamics of cosmological perturbations \cite{aan3}. First, because the scale factor of $\t{g}_{ab}$ never vanishes, the wave-lengths of modes never shrink to zero in LQC. Second, modes with wave lengths greater than the curvature radius of $\t{g}_{ab}$ at the bounce get excited in the LQC pre-inflationary dynamics and are not in the Bunch Davies (BD) vacuum at the onset of inflation. Among modes observable in the CMB, these turn out to be the longest wave length modes corresponding to angular scales $\ell \lesssim 30$. Thus, the LQC \emph{ultraviolet} modifications of the background FLRW dynamics lead to new observable effects in the \emph{infrared} regime of perturbations that are directly relevant to the large angular scale anomalies seen in the (WMAP and) Planck data.  This interplay between theory and observations is being investigated in detail from a number of different perspectives \cite{abrev}.

So far, this analysis of perturbations was carried out using sharply peaked states $\Psi$ for the background quantum geometry. As Fig.~\ref{3} shows, for these states the effective metric $\bar{g}_{ab}$ is so close to the quantum corrected dressed metric $\t{g}_{ab}$ that the difference between them is smaller than the numerical errors even in quite accurate simulations such as those performed in \cite{aan3}. While the use of sharply peaked states is well-motivated, it is of considerable interest to know whether the unforeseen and deep interplay between the ultraviolet and infrared regimes is robust. Does it persist if we replace the sharply peaked states with those in which the the relative dispersion in volume is non-negligible? A first step in systematically probing this issue is to analyze the difference between the mean value geometry $\gb_{ab}$ and the `dressed' metric $\t{g}_{ab}$ that governs the dynamics of perturbations, when both are determined by states $\Psi$ that have significant relative dispersions in volume. We will now report on the results of these calculations.
 
In order to compute the dressed scale factor, it is more convenient to work in the $(v,\,\phi)$ representation, rather than $(h,\phi)$, because one has to calculate the expectation values of a fractional power, $\widehat V^{4/3}$, of the volume operator $\widehat{V}$. In $(v,\,\phi)$ representation, the physical solutions to the quantum Hamiltonian constraint can be written as \cite{aps3}:
\be
 \Psi(v,\,\phi) = \int_0^\infty \!\!{\rm d}k\, \tilde{\Psi}(k)\, e_k(v) e^{i\alpha k (\phi-\phi_o)},
\ee
where $\phi_o$ is some initial time and $e_k(v)$ are the eigenfunctions of the  Hamiltonian operator. For our analysis we consider $\tilde{\Psi}(k)$ to be Gaussian wave-packets,
\be
  \tilde{\Psi}(k) = e^{\f{-(k-k_o)^2}{\sigma^2}},
\ee
and numerically compute the difference between the dressed and mean value geometries.

The results are shown in Figs.~\ref{3} and \ref{4}. Note that the  expression (\ref{tilde})  of `dressed' conformal time $\t\eta$ and the dressed scale factor $\t{a}$ in terms of $\Psi$ are quite complicated. Since they involve fractional and negative powers of the Hamiltonian $\h{H}_{o}$ and the 4/3rd power of the volume operator $\widehat{V}$, a priori there is no reason to expect that $\t\eta$ and $\t{a}$ would be well-approximated by the mean value quantities $\bar\eta$ and $\bar{a}$ which are computed simply using the expectation values of $\h{p}_{\phi}$ and $\h{V}|_{\phi}$. Yet we see that the dressed quantities are extremely well approximated by the mean value quantities within the accuracy of $0.02\%$ for scale factor and $0.04\%$ for the conformal time, if one considers a sharply peaked state having the late time relative volume dispersion $0.035$. Moreover, even when the relative dispersion in volume at late times is as high as $0.2$, the two scale factors approximate each other to within a 0.5\% accuracy. If one were interested only in 10\% accuracy, one can use the easy to calculate mean value metric $\gb_{ab}$ in place of the much more subtle $\t{g}_{ab}$ even when the relative dispersions in volume are greater than 1! 

These results provide a starting point for a systematic analysis of the robustness of pre-inflationary dynamics that will be reported in \cite{aag}. Interestingly, it turns out that, although the power spectrum does change if one uses states that fail to be sharply peaked, these modifications can be absorbed simply by changing the assignment of the phenomenological parameters, \emph{without} enlarging the parameter space underlying the phenomenological predictions of LQC.

\section{Discussion}
\label{s5}

In LQC, the standard effective description \cite{vt,asrev} has proved to be a powerful tool for understanding the physics of the bounce \cite{aps3,asrev}. As explained in sections \ref{s1} and \ref{s2}, the effective Friedmann and Raychaudhuri equations, (\ref{fried}) and (\ref{ray}), incorporate the key quantum corrections to Einstein's equations. They bring out the fact that quantum geometry effects lead to a new repulsive force. They show that the force is completely negligible in situations where general relativity is known to be successful. However, they also show that the repulsive force grows rapidly in the Planck regime, overwhelming the classical attraction and replacing the big bang with a big bounce. Thus, the effective equations make the physical origin of the quantum bounce transparent. They have also played a role in the investigations of pre-inflationary dynamics of cosmological perturbations. In particular, they have helped us understand why the ultraviolet modifications of the background FLRW dynamics can leave imprints on observable modes with longest wave lengths \cite{aan3,abrev}.

However, this highly successful effective description is tied to quantum states $\Psi$ which are sharply peaked in the principal Dirac observables $\h{p}_{\phi}$ and $\h{V}|_{\phi}$. In section \ref{s3} we extended that effective description to go beyond the sharply peaked states. Specifically, using the known expressions of the evolution of expectation values of $\h{p}_{\phi}$ and $\h{V}|_{\phi}$, we constructed a mean value metric $\gb_{ab}$ and obtained a generalization, (\ref{genfried}) and (\ref{genray}), of the effective Friedmann and Raychaudhuri equations, that governs the dynamics of $\gb_{ab}$. We also showed that the scalar field $\phi$ that underlies the construction of the mean field geometry automatically satisfies the Klein-Gordon equation with respect to the mean value metric $\gb_{ab}$. Thus, by focusing on expectation values, we obtained a consistent  generalization of the effective space-time description of the Einstein-scalar field system, now adapted to quantum geometry states $\Psi$ that need not be sharply peaked. Interestingly this space-time description continues to be internally consistent no matter how large the dispersions are. In section \ref{s4}, we used an interesting interplay between the area gap and the matter density at the bounce to show that Eqs. (\ref{genfried}) and (\ref{genray}) can be derived from a generalized effective Hamiltonian constraint. We then restricted ourselves to Gaussian states, but with possibly large relative dispersions in  $\h{V}|_{\phi}$ --i.e., the scale factor-- and showed that there is a simple relation, (\ref{rel}), between the density at the bounce and this relative dispersion which quantifies previous observations from numerical simulations \cite{dgs2,dgms}: greater the dispersion, lower is the density at the bounce. Finally, we compared the mean value metric $\gb_{ab}$ with the `dressed' quantum corrected metric $\t{g}_{ab}$ that is used to analyze cosmological perturbations in the Planck regime \cite{aan3,abrev}. We found that $\gb_{ab}$ can approximate $\t{g}_{ab}$ quite well even when the quantum state $\Psi$ is not sharply peaked. This last result provides a starting point for the investigation of how phenomenological predictions of LQC are affected if the quantum geometry state is not sharply peaked \cite{aag}. 

On the whole, the GEEs nicely complement the numerical calculations of the exact quantum evolution of widely spread states that have now become feasible, thanks to the development of the ``Chimera numerical scheme'' \cite{dgs1}. Since this numerical scheme solves the LQC equations exactly, it provides the evolution not only of the mean values of the principal Dirac observables but also of their fluctuations and higher order moments. These cannot be read-off from the GEEs. On the other hand, the GEEs provide a conceptual understanding of some of the key features of the exact numerical results. Furthermore, as we saw in section \ref{s4.3}, even in pre-inflationary dynamics where the dressed effective metric is sensitive to certain `higher moments' beyond the mean values of the principal Dirac observables, the GEEs provide a good approximation to the exact numerical results. Since the GEEs are \emph{ordinary} differential equations, computationally they are significantly easier to handle. Therefore, they provide an extremely useful tool to obtain a global understanding of the effect of changing the initial state at the bounce, which can then be followed by exact ``Chimera''  analysis when a more accurate treatment is warranted.

Finally, while summarizing the standard LQC framework \cite{asrev}, we took the opportunity in section \ref{s2} to slightly reformulate it to bring out more clearly the role of the area gap. Important formulas in the existing LQC literature generally feature both the area gap $\ag$ and the Barbero-Immirzi parameter $\gamma$, although one can be eliminated in favor of the other because they are proportional. This redundancy can cause occasional confusion while taking limits. While $\gamma$ features prominently in the construction of the mathematical framework, in the finished picture the new physical parameter of the \emph{quantum} theory is the area gap $\ag$. To emphasize this fact, throughout this paper we worked \emph{only with} $\ag$. This strategy makes the relation between LQC and the Wheeler DeWitt theory more transparent: The Wheeler-DeWitt equation results from the LQC quantum constraint (\ref{hc1}) by letting $\l$ --or equivalently, $\ag$-- go to zero. Similarly, classical Friedmann and Raychaudhuri equations are obtained as limits of (\ref{genfried}) and {\ref{genray}) when $\l$ goes to zero. One no longer has to carefully monitor what happens to the Barbero-Immirzi $\gamma$ parameter in various limits.

We will conclude with an example to illustrate why this shift of emphasis can be physically illuminating \cite{abrev}: it brings out a qualitative similarity between the quantum effects that lead to the singularity resolution in LQC and the physics of superconductivity. In the theory of superconductivity the fundamental microscopic parameter is the energy gap $\Delta_{\rm E}$, the energy needed to break apart the cooper pair of electrons in a given material. It determines the macroscopic behavior of that sample. In particular, the critical temperature is given by $T_{\rm crit} = ({\rm const})\,\, \Delta_{\rm E}$: as the the energy gap $\Delta_{\rm E}$ goes to zero in sample materials, the critical temperature goes to zero and we no longer have the novel phenomenon of superconductivity. Similarly, in LQC the microscopic parameter $\ag$ determines macroscopic observables such as $\rho_{\rm sup}$, the upper bound of the spectrum of the density operator $\h\rho$:\, $\rhosup = ({\rm const})/ {\Delta}_{o}^{3}$.\,\, If we were to send the area-gap $\ag$ to zero  --i.e., ignore the quantum nature of geometry underlying LQG-- \, $\rho_{\rm sup}$ would diverge, quantum effects would disappear, and we would be led back to the big bang singularity of GR.

\section*{Acknowledgments}
We would like to thank Parampreet Singh and especially Ivan Agullo for ongoing discussions. This work was supported by the NSF grant PHY-1505411 and the Eberly research funds of Penn state.

\begin{appendix}
\section{Calculations of expectation values and fluctuations}
\label{a1}

In this appendix we will briefly describe the soluble LQC model from \cite{acs} that is used in the main body of the paper, and provide expressions of various quantities used in section \ref{s4.2}.

\subsection{The physical Hilbert space}

Let us begin by recalling that the quantum Hamiltonian constraint given in (\ref{hc1}) can be rewritten as the 2-d flat space Klein-Gordon equation,
\be
\label{hckg} \partial_\phi^2 \Psi (x,~\phi) = \partial^2_x \Psi(x,~\phi),
\ee
by introducing
$$
 x = \f{1}{\sqrt{12\pi G}} \ln\,|\tan\f{\l h}{2}|\,.
$$
The group averaging procedure of LQG \cite{aps2} can then be used to construct the physical Hilbert space. The physical states are certain positive  frequency solutions to (\ref{hckg}). Every solution to (\ref{hckg}) can be written as a sum of left and right moving components:
\be
 \Psi(x,~\phi) = \Psi_{\rm L} (x_+) + \Psi_{\rm R} (x_-)\, ,
\ee
where $x_{\pm} = x\pm \phi$. In addition to being a solution to the quantum Hamiltonian constraint, the physical states have to satisfy the symmetry requirement, $\Psi(-x,\phi)=-\Psi(x,\phi)$, which captures the fact that the physical states remain invariant under the change of the orientation of the triad. Therefore, $\Psi(x,~\phi)$ has the following general form:
\be
  \Psi(x,~\phi) = \f{F(x_+)-F(x_-)}{\sqrt{2}},
\ee
where $F(x_\pm)$ is a positive frequency solution to (\ref{hckg}). The physical Hilbert space can be described in terms of the left moving solution $F(x_+)$ --or, equivalently right moving $F(x_-)$-- alone, as the entire information of the physical states is contained in $F$. Here, we choose to work with $F(x_+)$. The physical inner product can then be conveniently written solely in terms of $F(x_+)$ as:
\be
\label{inprod} \left(\Psi_1,~\Psi_2\right)_{\rm phys} = -2\,i
\int_{-\infty}^{+\infty} {\rm d}x\,
\overline{F}(x_+) \partial_{x_+} F(x_+).
\ee
In the following, for the simplicity of the notations we will drop the 
subscript from $x_+$ and denote it by $x$. 

For the computation of the expectation values of the physical observable it is convenient to write $F(x)$ as the following Fourier transform:
\be
 F(x) = \int_0^{\infty}{\rm d} k~ \tilde{F}(k)~ e^{-i\,k\,(\phi+x)},
\ee
where the integral is taken only over the positive $k$-axis, because $F(x)$ is a positive frequency solution, and evaluated at a constant $\phi$. Therefore the norm of a physical state can be written as an integral at a constant $\phi$ slice:
\be
\|\Psi\|^{2} = -2\,i \int_{-\infty} ^{+\infty} \!\!{\rm d}x\, \overline{F}(x)\, \partial_x F(x)
          = 2 \int_0^\infty\!\! {\rm d}k\, \overline{\tilde F}(k) k \tilde{F} (k).
\ee
%

\subsection{Expectation values:} 

Having the physical Hilbert space at hand, we can now calculate the expectation values of the principal Dirac observables in term of $F$
\cite{acs}. We have:

\ba
 \langle \hat p_{\phi} \rangle &= \f{2\hbar}{\|\Psi\|^{2}}
   \int_{-\infty}^{\infty} \left |\f{dF}{dx}\right|^2 \, dx, \qquad
 \langle \hat p^2_{\phi} \rangle &= \f{-2~i~\hbar}{\|\Psi\|^{2}} \int_{-\infty}^{\infty} \overline{\f{dF}{dx}}\, \f{d^2 F}{dx^2}\, dx, \\
 \langle \widehat V\rangle_{\phi} &= V_+\, e^{\alpha \phi} + V_-\, e^{-\alpha \phi}, \qquad   
 \langle \widehat V^2\rangle_{\phi} &= W_0 + W_+\, e^{2 \alpha \phi} + 
                                         W_-\, e^{- 2\alpha \phi},
\ea 
where $\alpha=\sqrt{12\pi G}$,\,\, and 
\ba
   V_\pm &=& \f{1}{ \| \Psi \|^{2}}\f{4\pi \gamma \lp^2 \lambda}{\alpha} 
           \int_{-\infty}^{\infty}  \left |\f{dF}{dx}\right|^2 e^{\mp \alpha x} \, dx, \\
   W_0 &=& \f{1}{ \| \Psi \|^{2}}\f{2 i \pi \gamma^2 \lp^4 \lambda^2}
               {3 G} \int_{-\infty}^{\infty}  \widetilde W(x)\, dx,  \\ 
   W_\pm &=& \f{1}{ \| \Psi \|^{2}}\f{i \pi \gamma^2 \lp^4 \lambda^2}
               {3 G} \int_{-\infty}^{\infty}  \widetilde W(x) e^{\mp 2\, \alpha\, x} dx ,
\ea
and 
\ba
  \widetilde W (x) &=& \overline{\f{d^2 F}{dx^2}}\, \f{dF}{dx} 
                     - \overline{\f{dF}{dx}}\, \f{d^2 F}{dx^2}.
\ea

\subsection {Relative volume dispersion:} 

Using the expressions for the expectation values of the volume observables we can compute the relative volume dispersion:
\ba
 \left(\f{\Delta V}{\overline V}\right)^2 &=& 
     \f{\langle \widehat V^2\rangle_{\phi} - \langle \widehat V\rangle_{\phi}^2}
         {\langle \widehat V\rangle_{\phi}^2} \\
        &=&\f{W_0 +(W_+-V_+^2) e^{2 \alpha \phi} + 
               (W_--V_-^2) e^{-2\alpha \phi} - 2 V_+V_-} 
             {V_+^2 e^{2 \alpha \phi} + V_-^2 e^{-2 \alpha \phi}+2V_+V_-}.
\ea
In order to find the relative dispersion far in future in the expanding branch
we take $\phi\rightarrow\infty$ limit in the above expression:
\be
  \left(\f{\Delta V}{\overline V}\right)_{\rm late\,time}^2 = \f{W_+}{V_+^2} -1,
\ee
similarly in the contracting branch:
\be
  \left(\f{\Delta V}{\overline V}\right)_{\rm early\,time}^2 = \f{W_-}{V_-^2} -1.
\ee

With these expressions at our disposal, one can compute the expectation value of the physical observables by specifying $\tilde F(k)$.

\subsection{Gaussian states}

In sections \ref{s4.2} we restricted our analysis to Gaussian states:
\be
  \tilde F(k) = \exp {-\f{(k-k_0)^2}{\sigma^2}}\,\,,
\ee
with the assumptions: $k_o \gg \sigma$ and $k_o\gg\alpha = \sqrt{12\pi G}$. Note that the physical states we have considered here are positive frequency solutions to the quantum Hamiltonian constraint, i.e. the integrals involved in the computation of norm and expectation values are restricted to the positive $k$-axis. The analytical expressions given in section \ref{s4.2} were obtained by integrating over the entire $k$-axis, i.e. from $-\infty$ to $\infty$. For the Gaussian states with large $k_o$ we are considering, this simplification introduces very small errors \cite{Corichi:2011rt}. In the following we provide estimates of these errors.
%
\begin{itemize}
 \item {\bf Norm:}  

      \be
        \| \Psi \|^2_{\rm phys} = \sqrt{2\pi} k_o\sigma  + {\rm Error_{\rm norm}}, 
      \ee

      where, ${\rm Error_{\rm norm}} = \f{1}{4} \sigma^2 e^{-2 k_o^2/\sigma^2}-
\sigma k_o \sqrt{\f{\pi}{2}} {\rm erfc}\left(\f{\sqrt{2}k_o}{\sigma}\right)$ is
approximation error in the norm \cite{Corichi:2011rt}. For states considered in
section \ref{s4.2} the numerical values for the approximation errors in the
norm are:
$\mathcal O(10^{-346})$ for $\sigma/k_{o} = 5\times 10^{-2}$ and
$\alpha/\sigma = 2.5\times 10^{-2}$ considered in (\ref{dvv1}), and 
$\mathcal O(10^{-23975})$ for $\sigma/k_{o} = 6 \times 10^{-3}$ and
$\alpha/\sigma =1.02$ considered in (\ref{dvv2})

 \item {\bf $\langle \hat p_\phi \rangle$:}
    \be
       \langle \hat p_\phi \rangle = \hbar k_o\left(1+\f{\sigma^2}
                                 {4 k_o^2}\right) + {\rm Error}_{p_\phi},
    \ee 
      where ${\rm Error}_{p_\phi}=\f{1}{8}\sigma\left(4
k_o\sigma e^{-2k_o^2/\sigma^2}-\sqrt{2\pi}(4k_o^2+\sigma^2){\rm
erfc}\left(\f{\sqrt{2}k_o}{\sigma}\right)\right)$ is the approximation error in
the computation of $\langle \hat p_\phi \rangle$. For the
Gaussian states considered in section \ref{s4.2} the approximation error is:
$\mathcal O(10^{-345})$ for $\sigma/k_{o} = 5\times 10^{-2}$ and
$\alpha/\sigma = 2.5\times 10^{-2}$ considered in (\ref{dvv1}), and
$\mathcal O(10^{-23977})$ for $\sigma/k_{o} = 6 \times 10^{-3}$ and
$\alpha/\sigma =1.02$ considered in (\ref{dvv2})
       
  \item {\bf $V_{\pm}$:}
     \be
       V_{\pm} = \f{\Delta_o^{3/2} \lp^3}{4\sqrt{3}\alpha}\, e^{\f{\alpha^2}
                 {2\sigma^2}} \left(4k_0 + \f{\sigma^2+\alpha^2}{k_0}\right)  +
                  {\rm Error}_{V_{\pm}}.
     \ee
     ${\rm Error}_{V_{\pm}} = \f{1}{16} \sigma e^{\f{-4k_o^2+\alpha^2}
                         {2\sigma^2}}
            \left|  -2 \sigma (2k_o+i \alpha)
e^{\alpha\f{4ik_o+\alpha}{2\sigma^2}} + e^{2k_o^2/\sigma^2} \sqrt{2\pi} (4
k_o^2+\alpha^2+\sigma^2) {\rm erfc}\left(\f{2k_o-i\alpha}{\sqrt{2}\sigma}\right)
\right| $, is the approximation error in the estimation of $V_\pm$. For the
Gaussian states considered in section \ref{s4.2} the approximation error is:
$\mathcal O(10^{-345})$ for $\sigma/k_{o} = 5\times 10^{-2}$ and
$\alpha/\sigma = 2.5\times 10^{-2}$ considered in (\ref{dvv1}), and
$\mathcal O(10^{-23974})$ for $\sigma/k_{o} = 6 \times 10^{-3}$ and
$\alpha/\sigma =1.02$ considered in (\ref{dvv2})

\item {$W_0$, $W_\pm$:} 
\ba
 W_0   &=& \f{\pi\Delta_o^3 \lp^6}{288 \pi G} \left(4k_0^2+3\sigma^2\right) + 
                     {\rm Error}_{W_0}\\
 W_\pm &=& \f{\pi\Delta_o\lp^6}{576 \pi G} \, e^{\f{2\alpha^2}{\sigma^2}} 
              \left(4k_0^2+4\alpha^2+3\sigma^2\right)+
                     {\rm Error}_{W_\pm}.
\ea
Similarly to the norm and $V_\pm$, the approximation errors in $W_0$ and 
$W_\pm$ are given in terms of error function. For the Gaussian states 
considered in section \ref{s4.2} the numerical values of the approximation errors in 
the estimation of $W_0$ and $W_\pm$ are: 
$\mathcal O(10^{-345})$ for $\sigma/k_{o} = 5\times 10^{-2}$ and
$\alpha/\sigma = 2.5\times 10^{-2}$ considered in (\ref{dvv1}), and
$\mathcal O(10^{-23975})$ for $\sigma/k_{o} = 6 \times 10^{-3}$ and
$\alpha/\sigma =1.02$ considered in (\ref{dvv2})

\end{itemize}

The above analysis shows that for large $k_o$ and $\sigma\ll k_o$, the
approximation errors in evaluating the integral over the entire $k$-axis are indeed extremely small. Therefore, for the Gaussian states of physical interest, the expectation values of the physical observables and their fluctuations for Gaussian states can be given, to an excellent approximation by:
\ba  
\langle \hat{V} \rangle|_{\phi} &=& \f{\Delta_o^{3/2} \lp^3}{4\sqrt{3}\alpha}\, e^{\f{\alpha^2} {2\sigma^2}} \left(4k_0 + 
\f{\sigma^2+\alpha^2}{k_0}\right) \cosh\left(\alpha \phi\right) \\
\langle \hat p_{\phi} \rangle &=& \hbar k_o \left(1+\f{\sigma^2}{4 k_o^2}\right)
\\
 \Delta p_{\phi} &=& \f{\hbar \sigma}{2} \sqrt{1-\f{\sigma^2}{4 k_o^2}} \\
  \left(\f{\Delta V}{\overline V}\right)_{\rm late\,time}^2 
          &=& \f{W_+}{V_+^2} -1 
          = e^{\alpha^2/\sigma^2} \f{\left(1+\f{3\sigma^2+4\alpha^2}{4k_0^2}\right)}{\left(1+\f{\sigma^2+\alpha^2}{4k_0^2}\right)^2} - 1. 
\ea
Similarly, the energy density at the bounce takes the from:
\be
 \bar{\rho}_{\rm B} = \f{\langle \hat p_\phi\rangle^2}{8V_+V_-}= \f{18\pi}{G^2\hbar\Delta_o^3}
\f{\left(1+\f{\sigma^2}{4k_0^2}\right)^2}
              {\left(1+\f{\sigma^2+\alpha^2}{4k_0^2}\right)^2}.
\ee
These are the expressions used in the main text.

\end{appendix}

\end{document}